\documentclass[usenatbib,twocolumn]{mn2e}
\usepackage{graphicx,verbatim,amsmath,amssymb,enumitem}
\usepackage[table]{xcolor} 
\begin{document}
\topmargin-1cm
\defcitealias{AJCF}{WTCB}

\title{Using large galaxy surveys to distinguish $z \simeq 0.5$
quiescent galaxy models}

\author[Cohn and White]{J.D. Cohn${}^1$ and Martin White${}^2$\\
${}^1$ Space Sciences Laboratory and Theoretical Astrophysics Center,
  University of California, Berkeley, CA 94720,\\
${}^2$ Department of Astronomy and Department of Physics,
  University of California, Berkeley, CA 94720\\}

\maketitle
\date{\today}

\begin{abstract}
  One of the most striking properties of galaxies is the bimodality in
  their star-formation rates.  A major puzzle is why any given galaxy is
  star-forming or quiescent, and a wide range of physical mechanisms have
  been proposed as solutions.  We consider how observations, such as might
  be available in upcoming large galaxy surveys, might distinguish different
  galaxy quenching scenarios.
  To do this, we combine an $N$-body simulation and multiple prescriptions from
  the literature to create several quiescent galaxy mock catalogues.
  Each prescription uses a different set of galaxy properties (such as history,
  environment, centrality) to assign individual simulation galaxies as
  quiescent.  We find how and how much the resulting quiescent galaxy distributions differ
  from each other, both intrinsically and observationally.  
  In addition to tracing observational consequences of different
  quenching mechanisms, our results indicate which sorts of quenching models might
  be most readily disentangled by upcoming observations and which
  combinations of observational quantities might provide the most
  discriminatory power.  

Our observational measures are auto, cross, and marked
correlation functions, projected density distributions, and group
multiplicity functions, which rely upon galaxy positions, stellar masses
 and of course quiescence.  Although degeneracies between models
are present for individual observations, using multiple
observations in concert allows us to distinguish between all ten models we
consider.   In addition to identifying intrinsic and observational 
consequences of
quiescence prescriptions and testing these quiescence models against
each other and observations, these methods can also be used to validate
colors (or other history and environment dependent properties) in simulated mock
catalogues.

\end{abstract}

\section{Introduction}

As time marches forward, collapsed objects in the universe grow hierarchically
via merging and accretion.  Gas accreting onto halos serves as the fuel for
star formation in the galaxies residing in dark matter halos.
Not all galaxies are forming stars at a significant rate today however, so a
central puzzle in galaxy formation is: why are some galaxies star-forming while
others are quiescent?
Several possible mechanisms to quench star-forming in galaxies have been
identified.
For instance, the gas may already be used up in stars, or might be
stripped from the galaxies, or prevented from accreting onto them,
or heated, and so on (see, e.g., the galaxy formation textbook by
\citealt{MoBosWhi10}).
Determining which of these processes are most significant for quenching
is a major challenge.  Due to the vast range of physical processes and
scales which can contribute to the quenching of star-formation
(e.g.~formation of stars, stellar feedback, and accretion onto and
outflows from black-holes) empirical constraints can play a vital role.

Here we consider ways in which large statistical galaxy surveys can be
used to distinguish between different models for galaxy quenching,
motivated by the tremendous impact of the Sloan Digital Sky Survey
\citep{York00} on studies of local galaxies, and other large surveys
either beginning or being planned.  We take such surveys to provide a
set of galaxy positions and stellar masses and whether or not the
star-formation in the galaxy is quenched.  We shall not consider here
detailed measurements of individual galaxies on smaller scales, such
as galaxy structural properties, properties such as winds, or AGN
signatures.  While these are indeed valuable, our focus is on
properties associated with large-scale structure as is available in
large surveys.

We expect that using even relatively little information per galaxy can
be highly informative when combined with large statistical samples,
because many proposed galaxy quenching mechanisms differ in their
dependence upon galaxy formation histories and environments.  These
differences lead to different overall statistical properties.  For
example, heating of a galaxy by an AGN may be more likely to occur
when the galaxy is in the core of a massive halo, while stripping of
gas could be more likely when a galaxy is moving through a very dense
environment (e.g.~as a satellite in a massive halo).  The two
resulting populations have different demographics.

In this work we create mock, quenched galaxy samples in $N$-body simulations,
based upon 10 different models and prescriptions drawn from or motivated by
models in the literature.
Each model assigns quiescence to individual mock galaxies based upon history
and/or environmental criteria, which we expect to be correlated with the
underlying physical (and baryonic) processes affecting star-formation.
We identify similarities and differences between these models, both intrinsic
and observable.
Our mock observations of these data help elucidate which future observations
might best disentangle such models, and how precise those observations need
to be.  How the observations interlock and constrain the models is also
valuable information for refining mock catalogs -- created via any method --
which are used in the design and analysis of large surveys.

We shall only consider a single redshift here ($z\simeq 0.5$); while
connecting galaxies across time is a powerful technique, it also
involves further assumptions.  At low redshifts ($z\simeq 0.1$) the
wealth of data and in depth analysis of rich data sets such as SDSS
(www.sdss.org) has provided many powerful constraints on environments
and other properties of quenched galaxies.  Some of our models
incorporate these insights.  By going back to $z\simeq 0.5$, we explore the
implications of these mechanisms for a significantly earlier cosmic
time, and one for which we expect large statistical samples in the
near future\footnote{For example from DES (www.darkenergysurvey.org),
  HSC (www.naoj.org/Projects/HSC/surveyplan.html), JPAS (j-pas.org),
  and LSST (www.lsst.org).}.  Note that $z\simeq 0.5$ is approximately
$5\,$Gyr ago, which is longer than the main-sequence lifetime of
$>1.4\,M_\odot$ stars.
Any galaxy less massive than $2\times 10^{10}\,M_\odot$ on the star forming
main sequence would at least double its stellar mass in this
period\footnote{Using Eq.~(2) of \citet{Lilly13}.}.
It is also around $z\simeq 0.5$ that we see a rapid rise in the number
density of intermediate-mass quiescent galaxies, making this a particularly
interesting time to study.

We also focus exclusively on a single galaxy property, quiescence.
Comparing measurements for one observed galaxy property, while fixing
the others, makes this work similar to the approach of
e.g.~\citet{HeaWat13,MasLinYos13} at $z\simeq 0.1$.
This forms a complement to more complex models, such as semi-analytic models,
which provide more information on each galaxy across cosmic time but also
more dependencies between inputs, assumptions and observations
\citep[see e.g.][for recent examples of fitting a suite of parameters
in such models to a suite of observations]{Lu11,NeiWei10,Lu13}. 

The outline of the paper is as follows.
In Section \ref{sec:simulations} we describe the $N$-body simulation and our
observation-based assignments of galaxy stellar mass and total quiescent
fraction.
In Section \ref{sec:models} we describe our 10 different prescriptions for
selecting quiescent galaxies based on histories and environments, and
in Section \ref{sec:comparison_int} and Section \ref{sec:comparison_obs}
we compare intrinsic and observational properties of the resulting
catalogues respectively.
Section \ref{sec:conclude} summarizes and concludes.
Throughout we use $\lg$ to denote $\log_{10}$, express stellar masses in
$M_\odot$ (with no factors of $h$), and halo masses in units of
$h^{-1}M_\odot$.  All distances and volumes are comoving and expressed in
$h^{-1}$Mpc or $h^{-3}\,{\rm Mpc}^3$.

\section{Simulations and stellar mass assignments} \label{sec:simulations}

In order to compare different models and mechanisms for quiescence, we
would like to have a sample of mock galaxies with environments,
formation histories and stellar masses which are close to what is seen
in observations.  To this end we associate mock galaxies with dark
matter subhalos (defined in more detail below) in an $N$-body
simulation.  Such associations are currently the standard tool for
analyzing galaxy surveys, as dark matter simulations of the cosmic web
are well converged \citep[e.g.][]{Hei08} and differences between dark
matter subhalo definitions and merger trees are becoming similarly well
characterized \citep{Oni12}.  The $N$-body simulation provides
subhalo properties such as positions, velocity, (dark matter) mass,
environment and history.  As dark matter is the dominant contribution
to the mass density in the Universe, these quantities are expected to
be close approximations\footnote{There are regimes where the
  approximation is known to break down, for instance on very small
  scales where baryons dominate; see \citet[e.g.][]{KuhVogAng12} for a
  recent review.  See also \citet{Wei08,Sim12,Daa13} for detailed comparisons
  of subhalo properties with and without baryonic effects.}  to their values when baryons and all interactions
are included (a comprehensive simulation of the latter is currently
infeasible).  We use the terms galaxy and subhalo interchangeably
henceforth.

We employ an $N$-body simulation performed in a periodic box of side
$250\,h^{-1}$Mpc.  This box has a similar volume to the main galaxy
survey of the SDSS, though without boundaries or gaps.  At $z\simeq
0.5$ it would subtend around $10^\circ$ on a side in the plane of the
sky.  The cosmology is of the $\Lambda$CDM family with
$(\Omega_m,\Omega_\Lambda,h,n,\sigma_8)=(0.274,0.726,0.7,0.95,0.8)$.
The simulation evolved $2048^3$ equal mass particles from initial
conditions generated at $z=150$ with second order Lagrangian
perturbation theory, using the code described in \citet{TreePM}.
Phase space data for all of the particles were dumped starting at
$z=10$ and for 45 times equally spaced in $\lg(1+z)$ down to $z=0$.  We
use the $z\simeq 0.5$ ($a=0.676$) output as our observation time.
Further details about this simulation can be found in \citet{WCS}.

For each output, halos are identified using the Friends of Friends (FoF)
algorithm \citep{DEFW}, with a linking length of $0.168$ times the mean
inter-particle spacing.
Halo masses quoted below are FoF halo masses unless otherwise specified.  
When halos merge, part of the smaller halo can survive as a self-bound
substructure within the larger host halo.  In such a situation we call the
core of the larger halo the ``central subhalo'' (or central) of the final
system and the core of the smaller halo which ``fell in'' a
``satellite subhalo'' (or satellite).
We reserve the term ``halo'' to refer to the parent FoF structure which
will host a central subhalo and possibly several satellite subhalos.
The subhalos are tracked as overdensities in phase space using the FoF6D
algorithm \citep{DieKuhMad06} as implemented in \citet{WCS}.
Merger trees and histories are calculated using the methods described in
\citet{WetCohWhi09,WetWhi10}.

As in \citet{WetWhi10} we only consider halos (and their descendant subhalos)
with masses above $10^{11.3} h^{-1} M_\odot$ to avoid resolution issues
(note that subhalos tracked using {\sc Subfind} \citep{SprYosWhi01} require
a higher minimum mass cut, as discussed in \citealt{GuoWhi12}).   
This cut results in 310,687 galaxies in our $z\simeq 0.5$ box or
$\bar{n}=7\times 10^{-3}\,{\rm Mpc}^{-3}$ (comoving). 
There are in addition 1, 18 and 110  (roughly cluster sized) halos with mass
above $10^{15}h^{-1}M_\odot$, $2\times 10^{14}h^{-1}M_\odot$ and
$10^{14}h^{-1}M_\odot$.  

\subsection{Stellar mass assignments}

Observationally, star formation is correlated with stellar mass, and
many of the models we investigate use as an input the stellar mass of
the galaxy.  We thus need a way to assign stellar masses to our mock
galaxies which results in the right demographics and 
environmental properties.
We assign stellar mass to each galaxy using subhalo abundance matching
\citep[e.g.][]{ValOst06,ConWecKra06} to the $z\simeq 0.5$ stellar mass
function of \citet{Dro09}.
We use this stellar mass function\footnote{Several other $z\simeq 0.5$
  stellar mass functions exist in the literature including
  \citet{Per08,Ilb10,Poz10,BehWecCon13a,Dav13,Kno13,Mar13,Mou13}.
  These are significantly different and such differences would propagate
  into the details of our modeling.
  Some sources of the differences are discussed in
  \citet{Mar09,Muz09,BehConWec10,Ilb10,Mou13,Tin13}.
  For example, \citet{Mou13} find small effects from changing stellar
  population synthesis models between
  FSPS \citep{ConGunWhi09,ConWhiGun10,ConGun10},
  \citet{BruCha03}, \citet{Mar05}, and
  Pegase \citep{FioRoc97,FioRoc99,LeB04},
  while \citet{BehConWec10} find a shift of $0.1\,$dex due to the
  choice of
  dust model.
  }
in part because several of our prescriptions start with the quiescent
fractions of central and satellite galaxies proposed by
\citet[][hereafter \citetalias{AJCF}]{AJCF}
based upon these data.

\begin{figure}
\begin{center}
\resizebox{3.5in}{!}{\includegraphics{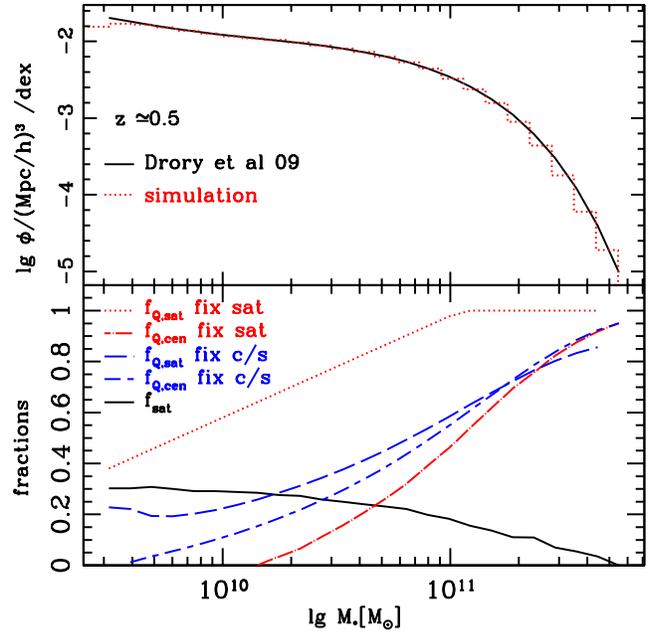}}
\end{center}
\caption{Top: The stellar mass function of \citet[][smooth curve]{Dro09}
and the histogram (dotted line) of stellar masses in our box.  This is used
as input to all 10 of our methods.
Bottom:  Satellite fraction at $z\simeq 0.5$ (solid black line), prescriptions for redshift
independent (dotted red line) and dependent (dashed blue line) quiescent
satellite fractions, and their corresponding quiescent central fractions.
These come from \citetalias{AJCF} and are used to construct the ``fix sat''
and ``fix c/s'' models described in Section \ref{sec:models}.
}
\label{fig:smf}
\end{figure}

We choose to abundance match stellar mass to the maximum mass in a
subhalo's history.  For a satellite subhalo, this maximum mass is
often the mass right before it becomes a satellite, sometimes known as
the infall mass.  (Using the maximum mass is along the lines
recommended by \citet{Red13}, who compared several possible abundance
matching proxies for stellar masses and luminosities at $z\simeq
0.05$, although the peak velocity was an even better proxy in their
case.)  We include a $0.16\,$dex scatter in stellar mass at fixed
maximum mass \citepalias[again motivated by the choices of][]{AJCF}
but we clip the scatter at $\pm 2\sigma$ to avoid outliers in the more
massive halos which are rare in our simulation box.  After assigning
$M_\star$, we keep only galaxies with $M_\star \geq
10^{9.5}\,M_\odot$.  Of the 310,687 mock galaxies in the box making
the halo resolution mass cut, 254,950 then remain
(of which 185,532 are central).
This galaxy sample is starting point for each of our catalogues below.
The stellar mass functions for \citet{Dro09} and for the simulation
box are shown in Fig.~\ref{fig:smf} where we see good agreement,
as expected.

\subsection{Quiescent fraction as a function of stellar mass}

\begin{figure}
\begin{center}
\resizebox{3.5in}{!}{\includegraphics{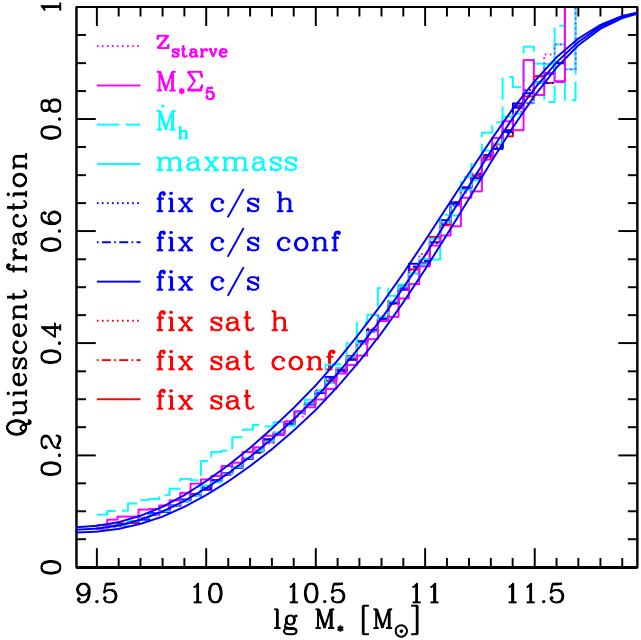}}
\end{center}
\caption{The quiescent fraction in our mock catalogues,
  all of which are tuned to the \citet{Dro09} observational fits
  (smooth blue lines).
  The central smooth blue line is the ratio of the best-fit quiescent stellar
  mass function to the quiescent plus active stellar mass functions (from the
  double Schechter fits in \citet{Dro09} Eq.~(1) and table 3).
  The two flanking blue curves are the $\pm 1\,\sigma$ variation of the
  4 normalizations ($\phi_b$ and $\phi_f$ for both quiescent and
  active galaxies) to give an indication of the range
  of observational uncertainty.
  The prescriptions based upon ``fix sat'', ``fix c/s'', ``maxmass,''
  and $z_{\rm starve}$ all match the \citet{Dro09} curve by construction,
  and so are all essentially degenerate.
  The red lines are for ``fix sat'' and the blue are for ``fix c/s''.
  The solid line is for random quiescent assignment, dot-dashed is our
  version of galactic conformity (``conf'') while dotted is satellite
  history (``h'').
  The solid cyan line is the ``maxmass'' model, while dashed cyan denotes
  $\dot{M}_h$; magenta solid is $M_\star\,\Sigma_5$ and magenta dotted is
  $z_{\rm starve}$.
}
\label{fig:redfrac}
\end{figure}

Each prescription we consider below assigns a subset of the galaxies
in our simulation box to be quiescent.  Matching observations of this
fraction is clearly a key property, and is built into the specification of several of our models.
For example, many of our models are taken from prescriptions of
\citetalias{AJCF}, which are in turn based upon the \citet{Dro09}
quiescent fraction as a function of stellar mass, $f_{Q,{\rm all}}(M_\star)$
(see Fig.~\ref{fig:redfrac}).   Others require a quiescent fraction as
function of stellar mass as input, again we choose that of \citet{Dro09}.
Specifically, we take $f_{Q,{\rm all}}(M_\star)$ as the number of quiescent
galaxies divided by the sum of quiescent and active galaxies in the double
Schechter function fits of Eq.~(1) and Table 3 of \citet{Dro09}.  

There is a large diversity in observed quiescent fractions as a function of
stellar mass.  The origins of some of these
differences are understood, in particular,
the \citet{Dro09} quiescent fractions are defined through SED fitting,
and thus differ from those employing other definitions
(e.g.~specific star formation rate, color cuts, morphology or some
combination).
\citet{Poz10} compare samples of quiescent galaxies selected by several
different definitions for a data set at $z\simeq 0.5$.
The galaxy samples differ physically and in number density, with color cut
based samples tending to include more galaxies because of dust.  
Two of our models are based upon non-SED based criteria for quiescence.
To take this additional (in part definitional) complexity out of our
comparisons, we modify our prescriptions, if needed, to improve agreement with
the SED based \citet{Dro09} $f_{Q,{\rm all}}(M_\star)$.
Unfortunately there is not always a unique modification to match to the SED
quiescent fraction, or matching it exactly is difficult.
We shall discuss these cases in detail below.  Our quiescent fraction
choice produces approximately 55,000-65,000 quiescent galaxies for
each catalogue, depending upon model.

\section{Prescriptions for assigning quiescent galaxies} \label{sec:models}

We now describe the different quiescent galaxy prescriptions we apply and the
10 resulting catalogues.

Our starting point is the $z\simeq 0.5$ box of mock galaxies with stellar
masses assigned as described above.  (By fixing this underlying galaxy
distribution for all 10
catalogues, any differences in quiescent galaxy catalogues
are due solely to the differences in assigning quiescence.)

The first and simplest model is motivated by a method by \citet{SkiShe09}
used at $z\simeq 0.1$ to assign colors to mock galaxies.
The method assumes we have a luminosity and central/satellite assignment for
each object, and that the color depends only on these properties.
Such a method gives good agreement with a number of observations at
$z\simeq 0.1$.  
In our case, we use each galaxy's stellar mass, $M_\star$, and whether it is a
satellite or central galaxy to determine the probability that it is quiescent. 
Individual galaxies are then marked, at random, using these probabilities.
With our assumptions, there is only one free function that we must specify to
determine the model, and we can take it to be the probability that a satellite
galaxy of stellar mass $M_\star$ is quiescent: $f_{Q,{\rm sat}}(M_\star)$.
The central quiescent fraction then is given by the requirement that we match
the overall quiescent fraction $f_{Q,{\rm all}}(M_\star)$ of \citet{Dro09}.
In more detail, and because conventions can  differ, we define
$f_{Q,{\rm sat}}(M_\star)$ as the fraction of satellites which are quiescent, i.e.~the
total number of satellite galaxies of a given $M_\star$ which are quiescent is
$f_{Q,{\rm sat}}(M_\star)\, f_{\rm sat}(M_\star)\, N_{\rm gal}(M_\star)$,
where $f_{\rm sat}$ is the fraction of $M_\star$ galaxies which are satellites
and $N_{\rm gal}(M_\star)$ is the total number of galaxies with stellar mass
$M_\star$.
The definition of $f_{Q,{\rm cen}}(M_\star)$ is analogous.

These functions have all been well constrained at $z\simeq 0.1$, but we expect
them to change with redshift.
Unfortunately, a precise measurement of these functions at higher redshift
is difficult as it requires a good group catalog and quiescence classification
over a cosmologically representative volume\footnote{Group
  catalogues, for small volumes (e.g., \citealt{Ger12,Geo13}), 
and estimates for associated quiescent fractions (e.g. \citealt{Kno13,Tin13}) are just beginning to be made, based upon a variety of quiescence definitions.}.
To overcome this, \citetalias{AJCF} suggest extrapolating from $z\simeq 0.1$ to
higher $z$ in two ways and using the difference as a measure of uncertainty
in the model.
As more observations become available we can use the measured
$f_{Q,{\rm sat}}(M_\star)$ rather than these extrapolations.
In the meantime, including both models helps us to understand the observational
signatures we see later.

\subsection{Fixed $f_{Q,{\rm sat}}$}

The first extrapolation to redshift $z=0.5$, which we refer to as
``fix sat'', 
assumes \citep{AJCF}
\begin{equation}
\begin{array}{ll}
  f^{\rm fix \; sat}_{Q,{\rm sat}}(M_\star,z=0.5) &\equiv
  f_{Q,{\rm sat}}(M_\star, z=0) \\
  &=
  -3.42 + 0.40\lg M_\star/M_\odot  \; .
\end{array}
\label{eq:fixsat}
\end{equation}
It is shown for our redshift in Fig.~\ref{fig:smf} as the dotted (red)
line.
For the stellar masses of interest to us, the overall quiescent fraction of
galaxies is observed to drop rapidly to increasing redshift, while the
satellite fraction evolves more modestly
\citep[see e.g.~table 1 of][for a recent compilation]{AJCF}.
If we hold $f_{Q,{\rm sat}}$ fixed, the decrease in $f_{Q,{\rm all}}$ must
be obtained by decreasing $f_{Q,{\rm cen}}$.
Thus, in order to match the observed evolution of the total quiescent
fraction, the fraction of central galaxies which are quiescent in this
prescription drops rapidly with increasing redshift.

To implement Eq.~(\ref{eq:fixsat}) and the related models described below,
we assign the specified quiescent fractions in 100 bins covering
$9.5\leq\lg M_\star/M_\odot < 12.06$.
If either $f_{Q,{\rm sat}}(M_\star)$ or $f_{Q,{\rm cen}}(M_\star)$
go outside the region $[0,1]$, we clip its value and choose the
other fraction to reach the desired $f_{Q,{\rm all}}(M_\star)$.
For our first prescription, we thus randomly assign galaxies to be quiescent
only requiring that Eq.~(\ref{eq:fixsat}) above and the corresponding
$f_{Q,{\rm cen}}(M_\star)$ are satisfied in each stellar mass bin.

We extended this prescription in 2 ways, to get 2 more catalogues.
The first extension is a version of ``galactic conformity''\footnote{Several
    definitions of conformity are in use, some of which not only affect
    galaxies sharing the same halo, but also galaxies within a larger region.
    Ours is most similar to that used by \citet{Wei06} and \citet{RosBru09}
    to describe correlation functions for SDSS at low redshift.
    See also recent work by \citet{Kau13,Phi13}.}.   
The quiescent centrals are taken to be the same as those in the ``fix sat''
case above.  Satellites of quiescent centrals are initially all taken to
be quiescent as well.  If the resulting number of quiescent satellites is too 
small to satisfy Eq.~(\ref{eq:fixsat}) in a given $M_\star$ bin, more
satellites are randomly assigned to be quiescent.
If the number of quiescent satellites is instead too large, quiescent
satellites are randomly changed to active to get agreement.
Both cases occur, for higher and lower $M_\star$ respectively.
(As we shall see later, in a similar model with more
quiescent centrals, many more halos have star-forming satellites but quiescent
centrals, while there are no cases of quiescent satellites with star-forming
centrals.  We shall explore the observational consequences of this 
below.)  
 On average, $\sim 70\%$ of
the satellites in high mass halos are quiescent in this variant, compared
to about 60\% for the (random) ``fix sat'' model.
We call this prescription ``fix sat-conf''.

We explored a stronger form of galactic conformity as well, which we did not
include in our family of catalogues.
With our fixed choice of quiescent centrals, inherited from the ``fix sat''
prescription, we assigned a rank to each halo.
Halos with quiescent centrals were ranked first, in order of descending
(total) richness.  Halos with star-forming centrals followed the centrally
quiescent halos but were ranked in ascending order of richness.
Each satellite galaxy then inherited the ranking of its parent halo.
Within a bin of stellar mass the first ranked
$f_{Q,{\rm sat}}(M_\star)N_{\rm sat}(M_\star)$
of the satellites were marked as quiescent, with the rest being star-forming.
This prescription resulted in no star forming galaxies in the richest halos
which have quiescent centrals (down to halo masses below
$10^{14} h^{-1}M_\odot$).
As this is in clear conflict with observations, we work with the less extreme
``fix sat-conf'' prescription described above.  

\begin{figure}
\begin{center}
\resizebox{3.5in}{!}{\includegraphics{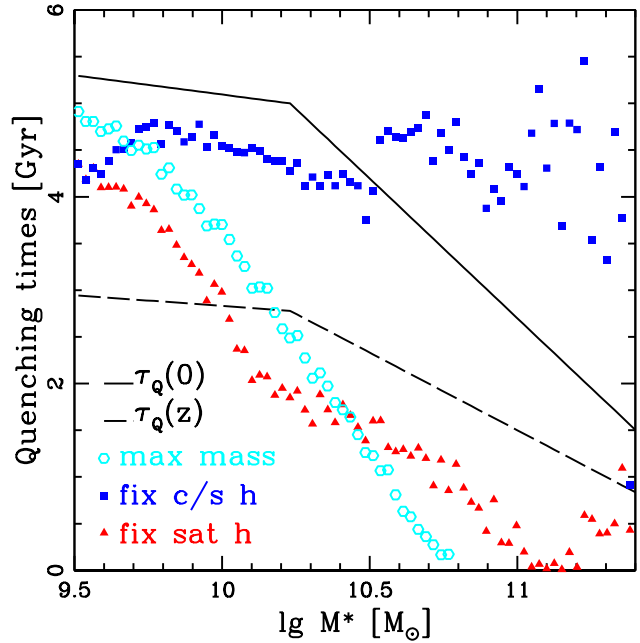}}
\end{center}
\caption{Quenching times found for ``fix sat-h'' (filled red triangles),
``fix c/s-h'' (filled blue squares) and ``maxmass'' (open cyan squares)
models as described in the text.  The solid black line shows the $z=0$
quenching times suggested for a model by \citetalias{AJCF} which is similar
to the ``maxmass'' model.  The (lower) dashed, black line gives a suggested
extrapolation to $z=0.5$ of the $z=0$ solid line, from \citet{TinWet10}.
At the highest stellar masses there are very few satellites, leading to the
larger scatter in quenching time.}
\label{fig:qtimes}
\end{figure}

The second extension based upon
Eq.~(\ref{eq:fixsat}), ``fix sat-h'', introduces a quenching ordering (and thus an implied time scale).
Central galaxies from the ``fix sat'' model are again left unchanged.
Each satellite is then ranked by its infall time.  To deal with the
discreteness in the output times from the simulation we take the infall time
to be random, uniformly distributed between the time step when the galaxy
was last a central and when it became a satellite.\footnote{This does add
scatter between infall times of galaxies which were part of the same halo
when falling into the larger halo, however we expect some scatter as
well in the quenching process.}
We take some satellites as quiescent upon infall with a probability based
on the extrapolations in \citetalias{AJCF} for $f_{Q,{\rm cen}}(M_\star,z)$
in this model as a function of redshift. (They give estimates for 4
stellar mass bins and we extend
the redshift scaling of their highest mass bin to apply to the most massive
satellites.)
We then take the remaining satellites in each stellar mass bin to be quiescent
in order of infall time (earliest first) to reach $f_{Q,{\rm sat}}(M_\star)$
in each stellar mass bin. 
The quenched satellite which fell in most recently of the latter set determines
a quenching time, shown as solid (red) triangles in Fig.~\ref{fig:qtimes}.
(At the highest stellar masses there are very few satellites, leading to the
larger scatter in quenching time.)  

These first three models differ in how they include environment or 
history, but the quiescent central galaxies are identical, as is the 
quiescent satellite fraction as a function of $M_\star$.  In
particular these models all have a relatively low central
quiescent fraction, as a direct consequence of Eq.~(\ref{eq:fixsat}).
We now consider the second extrapolation of $f_{Q,{\rm cen}}(M_*)$
from low redshift of \citetalias{AJCF}, to illustrate the impact of
this assumption.

\subsection{Fixed $f_{Q,{\rm cen}}/f_{Q,{\rm sat}}$}

An alternate way to evolve the quiescent satellite fraction, also suggested
by \citetalias{AJCF}, is to fix not $f_{Q,{\rm sat}}(M_\star)$ but the ratio of the
central and satellite quiescent fractions.  To match the observed drop in
$f_{Q,{\rm all}}$ towards higher $z$, both $f_{Q,{\rm sat}}$ and
$f_{Q,{\rm cen}}$ decrease.  However, since the drop is shared between the
two sources this prescription leads to a larger quiescent central fraction
than the assumption of Eq.~(\ref{eq:fixsat}).  Specifically we take
\begin{equation}
\begin{array}{lll}
  \frac{\displaystyle f^{\rm fix\; c/s}_{Q,{\rm cen}}}
       {\displaystyle f^{\rm fix\; c/s}_{Q,{\rm sat}}}(M_\star,z)
  &\equiv&
  \frac{\displaystyle f^{\rm fix \; c/s}_{Q,{\rm cen}}}
       {\displaystyle f^{\rm fix \; c/s}_{Q,{\rm sat}}}(M_\star,z=0) \\
  && \\
  &=& \frac{\displaystyle -6.12+0.64\lg M_\star/M_\odot }
           {\displaystyle -3.42+0.40\lg M_\star/M_\odot}
    \; \; . \\
\end{array}
\label{eq:fixcensat}
\end{equation}
Note, for $\lg M_\star/M_\odot<9.56$ the ratio is negative because the
quiescent central fraction at $z=0$ goes negative.
For these values of $M_\star$ we take all central galaxies to be active
($f_{Q,{\rm cen}}(M_\star)\equiv 0$) and set the quiescent satellite
fraction to obtain $f_{Q,{\rm all}}(M_\star)$ of \citet{Dro09}.

We can construct the same three variant models, random, conformity, and
satellite infall, using Eq.~(\ref{eq:fixcensat}) above
as a base instead of Eq.~(\ref{eq:fixsat}) as was done earlier.  Some
differences are immediately notable.
As just described, in comparison to the ``fix sat'' model, the ``fix
c/s'' model has more (or the same number of) quiescent centrals in all 
$M_\star$ bins at
$z\simeq 0.5$.   Due to the increase in number of quiescent centrals
relative to the ``fix sat'' models, the ``fix c/s-conf'' variation
has many more (roughly four times as many) satellites which are
active in halos with quiescent centrals, relative to the ``fix
sat-conf'' variation.
Furthermore, unlike the ``fix sat-conf'' case, there are no quiescent
satellites in halos with active centrals.
We shall see the implications of this below.
For the ``-h'' variation, we again find a the quenching time by assigning
galaxies with the earliest infall time as quiescent, but now up to the
fraction required by Eq.~(\ref{eq:fixcensat}).
This quenching time is much longer than that found for the ``fix sat''
models, hovering above 4 Gyr for almost all stellar masses, as can be
seen in Fig.~\ref{fig:qtimes}.  Satellites take longer to quench
because many more are quiescent upon infall, and because many fewer
are needed as quiescent overall, given Eq.~(\ref{eq:fixcensat}).

Further and more detailed comparisons between these prescriptions and
our other prescriptions are considered in
Sections \ref{sec:comparison_int} and \ref{sec:comparison_obs}.

\subsection{Other models}

Our four other prescriptions for when a galaxy becomes quiescent did not
use satellite and central quiescent fractions as a constraint.
All are based on models presented in the literature, although we have
modified or applied them to obtain the \citet{Dro09} quiescent fraction
$f_{Q,{\rm all}}(M_\star)$ as a function of stellar mass.

\subsubsection{Maximum mass}

In the ``maxmass'' prescription, satellites are quenched in order of the
earliest time since their maximum mass, up to the number needed to reach
the required $f_{Q,{\rm all}}(M_\star)$ in each $M_\star$ bin.
Similarly to our infall times for satellites, we smooth out the discreteness
of the output times in the $N$-body simulation by assigning the time of
maximum mass for each galaxy randomly and uniformly between the relevant
bracketing time steps (again this will erase some correlations between
galaxies sharing the same halo upon infall).
For satellites this maximum mass is also taken to be at or before infall time
-- we thus exclude mass gains through satellite merging, which can be
significant \citep[e.g.][]{Ang09,Sim09,WetCohWhi09}.
This implicitly assumes that mass gains due to merging after infall do not
induce further star formation, which seems reasonable.
Matching $f_{Q,{\rm all}}(M_\star)$ in each $M_\star$ bin then determines a
quenching time, which is comparable to that of the ``fix sat-h'' model
(open cyan circles and red filled triangles respectively in
Fig.~\ref{fig:qtimes}).

As the time of maximum mass is correlated with the time of first infall into
any larger halo \citep{AJCF13}, this model is similar to the
\citetalias{AJCF} model for quenching.  For comparison with their
model,  we also show their
quenching times in
Fig.~\ref{fig:qtimes}, both for $z=0$ (which they use, solid line) and one
estimate \citep{TinWet10} for the $z \simeq 0.5$ extension (dashed line).

\subsubsection{Halo growth}

The second of these prescriptions, which we label $\dot{M}_h$, is based on
a property of average star formation histories highlighted by
\citet{BehWecCon13a,BehWecCon13b}.
They calculated the average star formation rate as a function of stellar
mass from $0<z<8$, and compared with simulations to find an approximate
relation
\begin{equation}
  \frac{dM_\star}{dt}\approx \alpha(M_h) f_b \frac{dM_h}{dt}
\label{eq:sfdmh}
\end{equation}
i.e.~star formation rate proportional to baryon accretion rate, where the baryon
accretion rate is $f_b\, d M_h/dt$ and $f_b=0.17$ in their cosmology.
For $\alpha(M_h)$ independent of redshift, Eq.~(\ref{eq:sfdmh}) would imply a
redshift-independent $M_\star(M_h)$ relation.
While the $M_\star(M_h)$ relation does not evolve strongly, it is not
completely redshift independent  
\citep[e.g.][]{Yan12,AJCF,MosNaaWhi13,BehWecCon13b,WatCon13},
and thus this model can only be an approximation\footnote{Eq.~(\ref{eq:sfdmh})
was also calibrated to simulations with a certain step size which turns out to
be different from ours, and thus is approximate for this reason as well.  We
thank F.~van den Bosch for emphasizing this point to us.}.
We make further approximations by using Eq.~(\ref{eq:sfdmh}) to estimate the
sSFR for individual galaxies and then apply a cut on sSFR to classify a galaxy
as quiescent or star-forming, which is similar but not identical to an
SED-based classification (as mentioned above, and expanded upon below).

The halo mass change in Eq.~(\ref{eq:sfdmh}), $d M_h/dt$, is the change 
in $M_{\rm vir}$ halo mass.  We approximate the $M_{\rm vir}$ mass
gain by the FoF mass gain as the two mass definitions are close at this
redshift.
In addition, the change in observed stellar mass
(the observed star formation rate, of interest to us)
has a factor of 1.11 relative to the change in the true stellar mass given
above (from Eq.~(7) in \citet{BehWecCon13a}). 

Two additional assumptions are needed to identify which galaxies are
quiescent.  First of all, this approach does not indicate what to do
with satellite galaxies.
In addition, many central halos have their last significant mass gain
earlier than $z\simeq 0.5$.  Setting all galaxies in these categories
to have zero star formation well exceeds the \citet{Dro09} quiescent
fraction constraint.
We instead make an ansatz, again based upon the models in \citetalias{AJCF}:
we take galaxies to have a star formation rate from their most recent halo
mass gain (using $dM_h$ for that step and $dt$ for that step), but if that
time is before the present time step, we add a damping factor assuming that
the star formation is decaying from an earlier time.  (Again,
mass gains for satellites after infall, due to merging, are not included.)
Following \citetalias{AJCF}, we assume star formation actually started at $z=3$
and that its value at the time of most recent mass gain
(given by Eq.~\ref{eq:sfdmh})
has decayed by the time of observation.
Taking the star formation history for central galaxies
$\propto \Delta t \exp\left[-\Delta t/\tau_{\rm cen}\right]$,
with $\Delta t=t-t_{\rm form}$ and $t_{\rm form}=t(z=3)$,
\citetalias{AJCF} find $\tau_{\rm cen}$ within
$1.9-3.8\,$Gyr.
We take $\tau_{\rm cen}=3\,$Gyr. 
The star-formation rate at $z\simeq 0.5$ is then evolved from the value
given by Eq.~(\ref{eq:sfdmh}) at the time of most recent mass gain using
${\rm SFR}\propto \Delta t\, \exp\left[-\Delta t/\tau_{\rm cen}\right]$.
With only this prescription, however, the model does not have enough quenched
galaxies.
So in addition to increasing the star formation by using earlier time
steps, some quenching is needed to decrease the star formation by the
current time.
We thus augment the model by setting ${\rm sSFR}=10^{-13}\,{\rm yr}^{-1}$
for any central or satellite galaxy which had its more recent mass gain
more than a quenching time ago, assuming this quenching time depends upon
stellar mass, similar to our earlier models.
We have three examples of quenching times from our above
constructions: that of the ``fix sat-h'', ``fix c/s-h'' and
``maxmass'' models, in Fig.~\ref{fig:qtimes}.  We chose to use the
``fix sat-h'' quenching times as they gave the closest agreement to
the desired quiescent fraction $f_{Q,{\rm all}}(M_*)$ when combined
with the sSFR cut below.  (This allows
some interesting model comparisons below, as for``fix sat-h'' model
the times were only defined by and used for the satellites, with
centrals were assigned randomly; and here the times are used for all
galaxies.)  The quenching time is responsible for the quiescence of
about 1/5 of the quiescent centrals and about 4/5 of the quiescent
satellites (these latter overlap with the ``fix sat-h'' satellites as
a result of using that prescription's time scale).

The last piece needed to get some estimate of quiescence is a
comparison of this sSFR estimate to a quiescence classification based
on SEDs.  The sSFR assignments in \citetalias{AJCF} suggest a
cut\footnote{Note that the bimodality of star formation rates seen in
  observations (and in \citetalias{AJCF}) is not strong in our
  prescription, as more than half of the quiescent galaxies are such
  because their mass gains are before $\tau_{Q}(M_\star)$.
  The remainder of the quiescent galaxies are just the low end tail of
  the sSFR distribution.  There is a similar issue in \citet{MutCroPoo13}.}
between active and quiescent galaxies at sSFR$=10^{-11}{\rm yr}^{-1}$.
A comparison of SED and sSFR cuts for galaxies at similar redshifts is
found in fig.~1 of \citet{Poz10}.
Taking those numbers at face value, a sSFR cut of $10^{-11}{\rm
  yr}^{-1}$ does not include any active galaxies by the SED
definition, but also neglects some quiescent ones.
We change the maximum specific star formation rate for quiescent galaxies
to $3 \times 10^{-11}{\rm yr}^{-1}$, which in \citet{Poz10} would allow 
some admixture of active SED galaxies but also includes more quiescent
SED galaxies. The resulting galaxy sample gives better agreement numerically with the SED
determined quiescent fraction as a function of $M_\star$ in \citet{Dro09}.

Finally, we note that we do not include any scatter in the sSFR, taking it
directly from Eq.~(\ref{eq:sfdmh}), and we do not self-consistently integrate
$dM_\star/dt$ over time, instead applying it instantaneously at $z\simeq 0.5$.
A forward integration of Eq.~(\ref{eq:sfdmh}) may couple $M_\star$ and
$dM_\star/dt$ more closely, while inclusion of scatter would obviously weaken
such a correlation.

\subsubsection{Stellar mass and density}

Our third additional prescription, which we label ``$M_\star\,\Sigma_5$'', is based on
\citet[][see also \citealt{Kov13}]{Pen10}.
These authors found that the red and blue fractions of the $z$COSMOS sample
could be described by the product of two factors, depending upon
projected density $\Sigma_5$ and stellar mass,
\begin{equation}
\begin{array}{lll}
  f_{\rm red}(M_\star,\Sigma_5) &=&
  \epsilon_{\Sigma_5}+\epsilon_{M_\star}-\epsilon_{\Sigma_5} \epsilon_{M_\star}
  \\
  \epsilon_{\Sigma_5} &=& 1 - \exp[\left(-\Sigma_5/p_1\right)^{p_2}] \\
  \epsilon_{M_\star} &=& 1 - \exp[\left(-M_\star/p_3\right)^{p_4}] \; .\\
\end{array}
\label{eq:pengstart}
\end{equation}
with $(p_1,p_2,p_3,p_4)$ functions of redshift.
Here $\Sigma_5$ is the projected density, defined by fifth nearest neighbor
in a redshift cylinder of $\pm 1000\,{\rm km}\,{\rm s}^{-1}$, including
galaxies down to $M_{B,AB}\leq -19.8$
\citep[for $z\simeq 0.5$,][]{Kov10,Pen10}.

This model is particularly interesting for our purposes because it does not
specifically refer to halo mass or to central/satellite designation.  However
we need to modify it slightly.
Although we have $M_\star$ for each galaxy, the projected density assignment
as used by \citet{Pen10} relies upon a sample selected with $M_B$, and $M_B$
depends upon color, which is related to quiescence, which we are trying to find.
In addition, the parameters $p_i$ in Eq.~(\ref{eq:pengstart}) 
are tuned to separate galaxies into red
and blue (i.e. using observations based upon a color cut), rather than quiescent
and active.   We make two modifications as a result.

Instead of using an $M_B$-limited sample to define projected density, we use an
$M_\star$-limited sample.  We choose the minimum $M_\star$ to give the same
galaxy number density as the number of galaxies passing the $M_B$ cut of
\citet{Pen10}.
At $z\simeq 0.5$, the $M_B$ limit is $M_{B,AB}\leq -19.8$.
This corresponds to a density of $10^{-2}\,h^3\,{\rm Mpc}^{-3}$
using the COMBO-17, DEEP2 and  VVDS $B$-band luminosity functions quoted in
\citet{Fab07}.  Matching this number density fixes our threshold
$M_\star\simeq 8\times 10^9M_\odot$.
The projected density $\Sigma_5$  is then the inverse square of the distance to the
$5^{\rm th}$ nearest neighbor within the cylinder divided by its average value
for a sample of random positions within the box.

Using $\Sigma_5$ and $M_\star$ in Eq.~\ref{eq:pengstart} gives
each galaxy in our box a probability of being red, defined by the color cut
chosen in \citet{Pen10}.
The resulting $f_{Q,{\rm all}}(M_\star)$ well exceeds that of
\citet{Dro09}.  This is not surprising.  
 As mentioned
earlier, color cuts tend to classify more galaxies as quiescent (red)
than an SED (our comparison \citet{Dro09} sample) or sSFR based cut.\footnote{See also \citet{Tin13} for
  discussion of different quiescent fractions in these models.}  Our
$\Sigma_5$ assignment based upon $M_\star$ rather than $M_B$ might be
also a factor, however, this red fraction in our box is a good match to
that in \citet{Kno13}, including central and satellite red fractions
as a function of $M\star$.  (For this test we also use the
\citet{Kno13} stellar mass function for consistency.)

Our prescription is thus to keep the form of the \citet{Pen10} model,
but to modify the parameters to match the $f_{Q,{\rm all}}(M_\star)$
of \citet{Dro09}.  As \citet{Pen10} tuned their parameters using the measured
distributions of quiescence as a function of both $\Sigma_5$ and $M_\star$,
while we only have $f_{Q,{\rm all}}(M_\star)$, our modifications cannot be
unique.  (Some degeneracy is expected as there is some correlation of high
$M_\star$ with high $\Sigma_5$.)  We found 
$(p_1,p_2,p_3,p_4) = (202,1.71,1.56\times 10^{11},0.69)$ fit 
our fiducial $f_{Q,{\rm all}}(M_\star,z=0.5)$ quite well,
which corresponds to multiplying the \citet{Pen10} default values
(from their table 2, averaged for two redshift bins) by
$(3.2,2.5,2.5,1.1)$.

\subsubsection{Starvation}

Our last prescription is based upon a proposal by \citet{HeaWat13},
see also \citet{ZenHeavdb13}, of ``age matching''.
These authors assign $r$-band luminosities to a simulation similar to ours
using subhalo abundance matching, and then order galaxies in luminosity bins
according to a redshift $z_{\rm starve}$ (roughly a quenching time).
Galaxies with the earliest $z_{\rm starve}$ are taken to be quiescent,
up to the total number required (for us set by \citealt{Dro09}).
\citet{HeaWat13} define $z_{\rm starve}$ as the maximum (earliest) of
three possible redshifts: 
\begin{description}[leftmargin=*]
\item -when the host halo equals or exceeds $10^{12}\,h^{-1}M_\odot$.
\item -when the galaxy becomes a satellite.
\item -when the host halo growth rate slows down.
\end{description}
Each of these redshifts is taken to be zero if it never occurs.
\citet{HeaWat13} use $M_{\rm vir}$ in the first condition, while we shall
instead use $M_{\rm FoF}$.  The two definitions are close at the redshifts
of interest.
The halo growth rate is taken from \citet{Wec02}.
While \citet{HeaWat13} use a relation between concentration and halo growth
rate, we instead use the definition of \citet{Wec02}:
$d\log M_{\rm halo}/d\log a\equiv \Delta\log M_{\rm FoF}/\Delta\log a <2$. 
We search through the time steps in the simulation to find the earliest time
that the halo growth rate dropped below, and subsequently stayed below,
the threshold growth rate.

As in our previous models, we bin on $M_\star$ (rather than $r$-band luminosity).
We assign quiescent galaxies based upon the galaxies with the highest
$z_{\rm starve}$ in each of 100 stellar mass bins.
The last condition (slow down in growth) accounts for slightly more than 60
per cent of the $z_{\rm starve}$, with the remainder of the assignments
split approximately equally between the other two conditions.

A fifth additional model which we implemented, but which we do not
include below, is that of \citet{Lu13}.
These authors assign star formation rate based upon halo mass, infall time
for satellites, and stellar mass.  
Using their preferred model we found quiescent fractions too large to
be consistent with \citet{Dro09}.  In particular their model had a very short
satellite quenching time, making essentially all satellites quenched.
While there are combinations of parameters in their prescription which give
longer quenching times, without repeating their full likelihood analysis
we could not identify combinations which matched the our required quiescent fraction and the
other constraints they imposed.\footnote{We thank Z. Lu for
  discussions of their work.}

\subsection{Summary of the models}

\begin{table*}
\centering
\begin{tabular}{l|c|c|c|c|c|}
Model & \shortstack{cen or \\ sat} & \shortstack{sat infall \\ time} &
$M_h (t)$ & \shortstack{nearby \\ galaxies}&$\dot{M}_h$\\ \hline
\shortstack{``fix sat'' \\``fix c/s'' }
 &x & & & &  \\ \hline
\shortstack{``fix sat-conf'' \\ ``fix c/s-conf''} &x & & &x & \\ \hline
\shortstack{``fix sat-h'' \\ ``fix c/s-h''}
 &x &x & & &  \\ \hline
``maxmass''   &x &(x) &x & &  \\ \hline
$\dot{M}_h$&x&x&& &x\\ \hline
$M_\star\Sigma_5$& & & &x & \\ \hline
$z_{\rm starve}$&x &x&x& &x\\ \hline
\end{tabular}
\caption{Summary of which physical properties were used to assign quiescence
for each of the galaxy catalogues.  Some catalogues assign quiescence only
using whether a galaxy is a central or satellite, in contrast, the
$M_\star\,\Sigma_5$ catalog does not refer to host halos at all.
Only one model ($z_{\rm starve}$) depends upon host halo mass explicitly.
Several models depend upon infall time for satellites, the ``maxmass''
model only does because mass gain by definition does not happen after infall.}
\label{tab:pres}
\end{table*}

The quiescence criteria above result in 10 galaxy catalogues, differing
only by which galaxies are marked as quiescent.
The next step is to compare measurements on all these catalogues, however
it is first useful to summarize how the prescriptions differ.
A comparison of the criteria used in determining quiescence
(central or satellite in halo, infall time, halo mass change, etc.)
is given in Table \ref{tab:pres}.
In words, the criteria for quiescence can be briefly described as:
\begin{description}[leftmargin=*]
\item[``fix sat'' and ``fix c/s''] set quiescent satellite and central
fractions using Eqs.~(\ref{eq:fixsat}, \ref{eq:fixcensat}) respectively and
$f_{Q,{\rm all}}(M_*)$, and
then assign quiescence randomly 
\item[``fix sat-conf'' and ``fix c/s-conf''] as above but put quiescent satellites
preferentially in halos with quiescent centrals (the rest are random)
\item[``fix sat-h'' and ``fix c/s-h''] randomly assign some satellites as
quiescent at infall using proposed higher redshift behavior, assign
rest of quiescent satellites in order of infall time to reach the desired quiescent fraction 
\item[``maxmass''] assign galaxies which reach their maximum mass the
earliest as quiescent, up to the total quiescent fraction desired
\item[$\mathbf{\dot{M}_h}$] take sSFR from most recent halo mass gain;
also assign galaxies as quiescent if the most recent halo mass gain is before
time scale found in ``fix sat-h'' 
\item[$\mathbf{M_\star\,\Sigma_5}$] quiescent probability from the
stellar and (projected) local density using Eq.~(\ref{eq:pengstart}).
\item[$\mathbf{z_{\rm starve}}$] ranks galaxies by earliest of three possible
starvation times; assigns quiescence to galaxies in order, up to the number
needed.
\end{description}

\section{Comparisons: intrinsic} \label{sec:comparison_int}

With these catalogs in hand we wish to see in what ways the different
prescriptions lead to different populations of quiescent galaxies.
We begin with a discussion of the intrinsic properties, before turning
to the observational consequences in Section \ref{sec:comparison_obs}.

Two intrinsic properties are imposed on the catalogues by construction.
The first is the stellar
mass function, shown in Fig.~\ref{fig:smf} and implemented identically
in all the catalogues.  The second, the quiescent
fraction of all galaxies as a function of stellar mass,
$f_{Q,{\rm all}}(M_\star)$, was used implicitly or explicitly in the
catalogue construction as well
(and used to exclude the model of \citealt{Lu13} which did not provide a
 good match to this function).
In Fig.~\ref{fig:redfrac}, the observations of \citet{Dro09} are shown
along with the measurements from the models.
While the models generally reproduce $f_{Q,{\rm all}}(M_\star)$ quite well,
the match is not perfect for all of them.
In particular, the $M_\star\,\Sigma_5$ and $\dot{M}_h$ prescriptions
have slight excesses at low $M_\star$, and the $\dot{M}_h$ model is
also high at high $M_\star$, although in-between it matches the others
fairly well.

We now consider measurement other than these two intentionally degenerate
properties to highlight differences between the catalogues.

\subsection{Central and Satellite Quiescent Fractions}

\begin{figure*}
\begin{center}
\resizebox{3in}{!}{\includegraphics{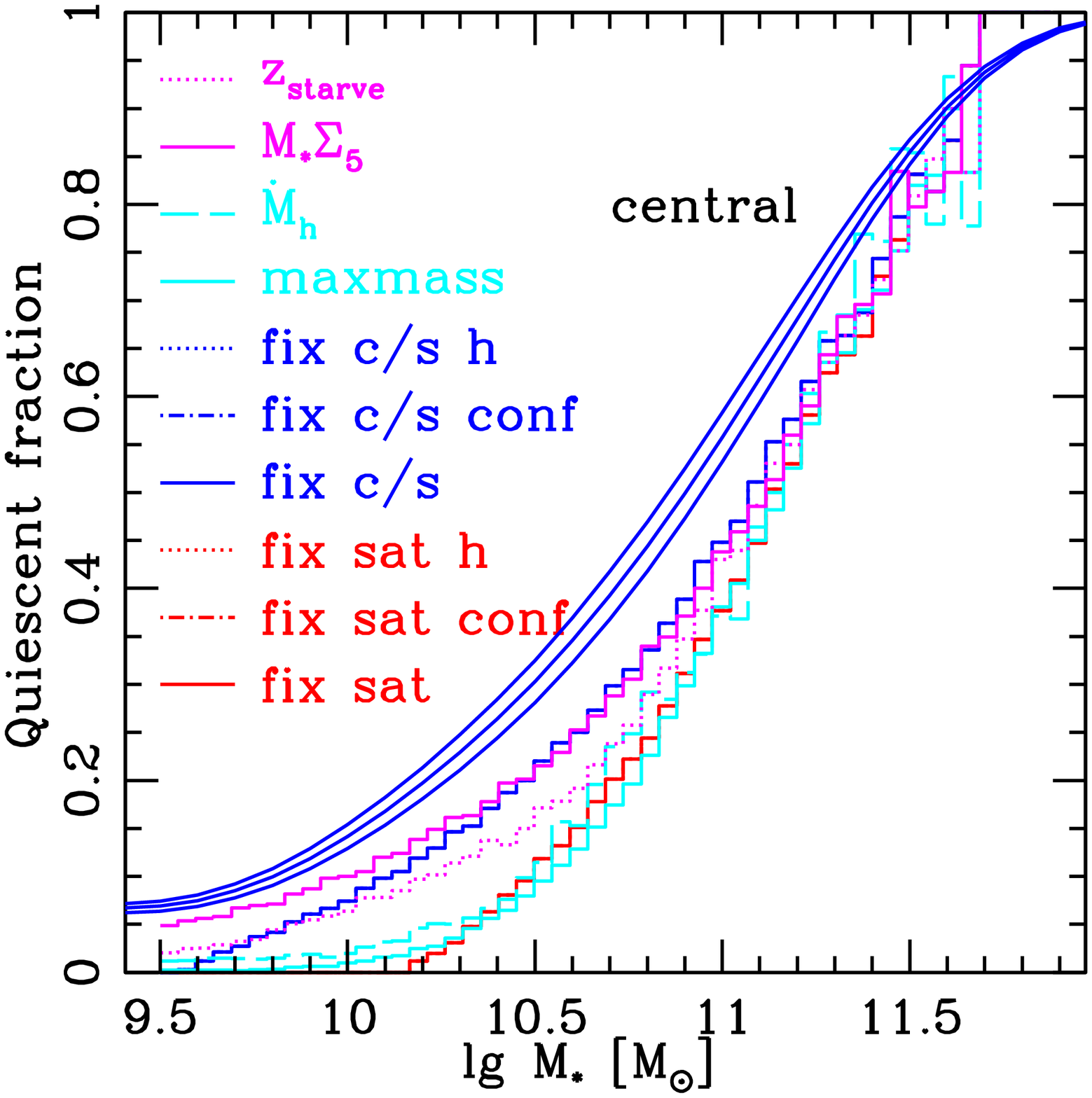}}
\resizebox{3in}{!}{\includegraphics{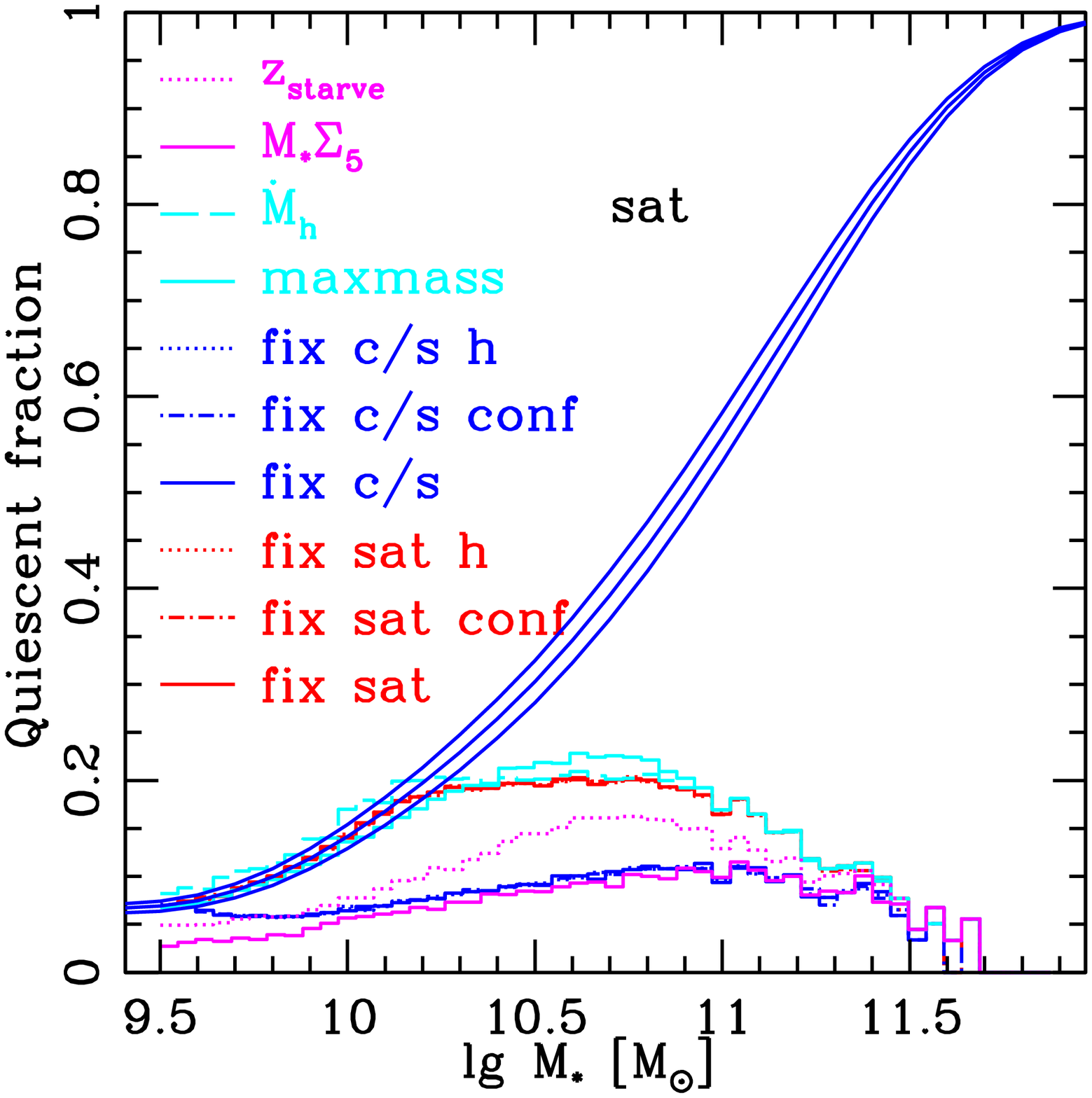}}
\end{center}
\caption{Contributions to the total quiescent fraction from central (left)
  and satellite (right) quiescent galaxies, as a function of stellar mass.
  Line types for each of the catalogues are as in Fig.~\ref{fig:redfrac},
  as are the smooth blue lines representing an observational range for the
  total quiescent fraction.
  The prescriptions roughly separate into two groups for both central and
  satellite galaxies, with the $z_{\rm starve}$ prescription lying in-between.
  The ``fix c/s'' prescription lies in the group with more quiescent central
  galaxies as a function of $M_\star$, while the ``fix sat'' prescription is
  in the group with fewer quiescent central galaxies.
  (The trend is reversed for the satellite quiescent galaxies, as the sum
  is fixed).  }
\label{fig:satcenred}
\end{figure*}

The quiescent fraction of all galaxies as a function of stellar mass,
$f_{Q,{\rm all}}(M_*)$ above,
can be decomposed into central and satellite contributions.
These are challenging to obtain observationally, which is in part the reason why
\citetalias{AJCF} propose two different $z\simeq 0.5$ quiescent satellite and
central decompositions (the starting points for our ``fix sat'' and ``fix c/s''
models).
However, understanding the way in which these functions differ between the
models will help us to understand the observational trends in the next section,
and which observations bear most directly upon this separation.

The breakdowns into quiescent central ($f_{Q,{\rm cen}}\,f_{\rm cen}$)
and quiescent satellite ($f_{Q,{\rm sat}}\,f_{\rm sat}$) galaxies for our
catalogues are shown in Fig.~\ref{fig:satcenred}.
For context we also reproduce the total quiescent galaxy fraction of
\citet{Dro09} from Fig.~\ref{fig:redfrac}.
The solid and dot-dashed lines for the ``fix sat'' (red) and ``fix c/s'' (blue)
models show the ``fix c/s'' model almost always has the same number or more
quiescent central galaxies than the ``fix sat'' model, as we have remarked
before.  This is a direct consequence of the assumed redshift dependence of
$f_{Q,{\rm sat}}$.  If the decrease in quenched fraction with increasing
redshift is borne entirely by the central galaxies, rather than being shared
equally by the central and satellite galaxies, then $f_{Q,{\rm cen}}$ is very
small.

Note that the models separate into two groups, with the $z_{\rm starve}$
prescription lying in-between.  The trends in the central, quiescent galaxies
are mirrored for the quiescent satellites as the sum is fixed.
The models with conformity and the ``-h'' models have the same
fractions as their parent models, because those models simply
``shuffle'' the quiescent or satellite galaxies into different halos.

Three of the four models which do not specify the satellite and central
quiescent fractions directly have strong similarities to the ``fix sat'' or
``fix c/s'' models.  In part this is because the quenching
prescriptions lead to 
overlap of the quenched populations.
The $\dot{M}_h$ model uses the quenching time after satellite infall of the ``fix sat-h''
model, 
$\sim$90 per cent of the $\dot{M}_h$ model quenched satellites
coincide with 80 per
cent of the ``fix sat-h'' quenched satellites.
The ``maxmass'' model also has
$\sim$90 per cent of its satellites in common with
the ``fix sat-h'' model (and a similar percentage in common with the
$\dot{M}_h$ satellites).
These models have much less overlap in their quiescent centrals:
quiescent centrals are randomly chosen for ``fix sat-h'', while 
those for $\dot{M}_h$ are quiescent because recent mass gain was too
small or too long ago, and for  ``maxmass'' because largest mass was
too long ago.  About 2/3 of the ``maxmass'' quiescent centrals are
also
$\dot{M}_h$ centrals, but only about 1/3 of $\dot{M}_h$ and
``maxmass'' centrals overlap with  ``fix sat-h'' quiescent centrals.
As about three times as many central galaxies as satellite galaxies
differ between
the ``maxmass'' and $\dot{M}_h$ models relative to the ``fix sat''
models, central galaxy population differences might be
the source of observational differences between the
``maxmass'' and $\dot{M}_h$ models and the ``fix sat'' model.

At lower $M_\star$ the satellite fraction is relatively large, and satellites
can saturate the required red fraction.  Since the halos of central galaxies
tend to gain mass more rapidly than those of satellites (whose last mass
gain was prior to their infall) the $\dot{M}_h$ and ``maxmass'' models
predict a very small fraction of quiescent, low $M_\star$ centrals.

By contrast, 
the $M_\star\,\Sigma_5$ model lies close to the initial ``fix c/s'' fractions.
This model has more low-$M_\star$, quiescent centrals than most of the
others.  In practice, 
the model makes little distinction between centrals and satellites, 
as the parameters $(p_1,p_2,p_3,p_4)$ which we found to fit the
$f_{Q,{\rm all}}(M_\star)$ constraint imply
$f_{Q,{\rm sat}}(M_\star) \approx f_{Q,{\rm cen}}(M_\star)$.  
These fractions are also fairly close for the  ``fix c/s'' models in
Fig.~\ref{fig:smf}.
If the satellite and central quiescent fractions are close, more central
galaxies will be quiescent because they outnumber satellites at any given
$M_\star$. 
(This satellite/central blindness is not necessarily part of the 
$M_\star\,\Sigma_5$ form for assigning quiescence.
At fixed stellar mass, there is a $\Sigma_5$ difference on average
between satellites and centrals, and parameters such as those in
\citet{Pen10} can lead to a large difference between
$f_{Q,{\rm sat}}(M_\star)$ and $f_{Q,{\rm cen}}(M_\star)$.) 
But by  sampling a wide range of parameters, it seems that getting
$f_{Q,{\rm all}}(M_\star)$ low enough at low $M_\star$ requires raising
$p_1$ at low $M_\star$, thus taking $f_{Q,{\rm sat}}(M_\star)$ towards
$f_{Q,{\rm cen}}(M_\star)$ as we found.)


We now turn from quiescent satellite and central stellar mass functions
to the question of which halos these satellites and central galaxies occupy.

\subsection{Quiescent Galaxy Halo Occupation}

The manner in which the low- and high-mass quiescent central and satellite
galaxies are distributed among halos differs significantly between the
models.  In this section we ask:
``in which types of halos are quiescent galaxies found?''

\begin{figure*}
\begin{center}
\resizebox{3.in}{!}{\includegraphics{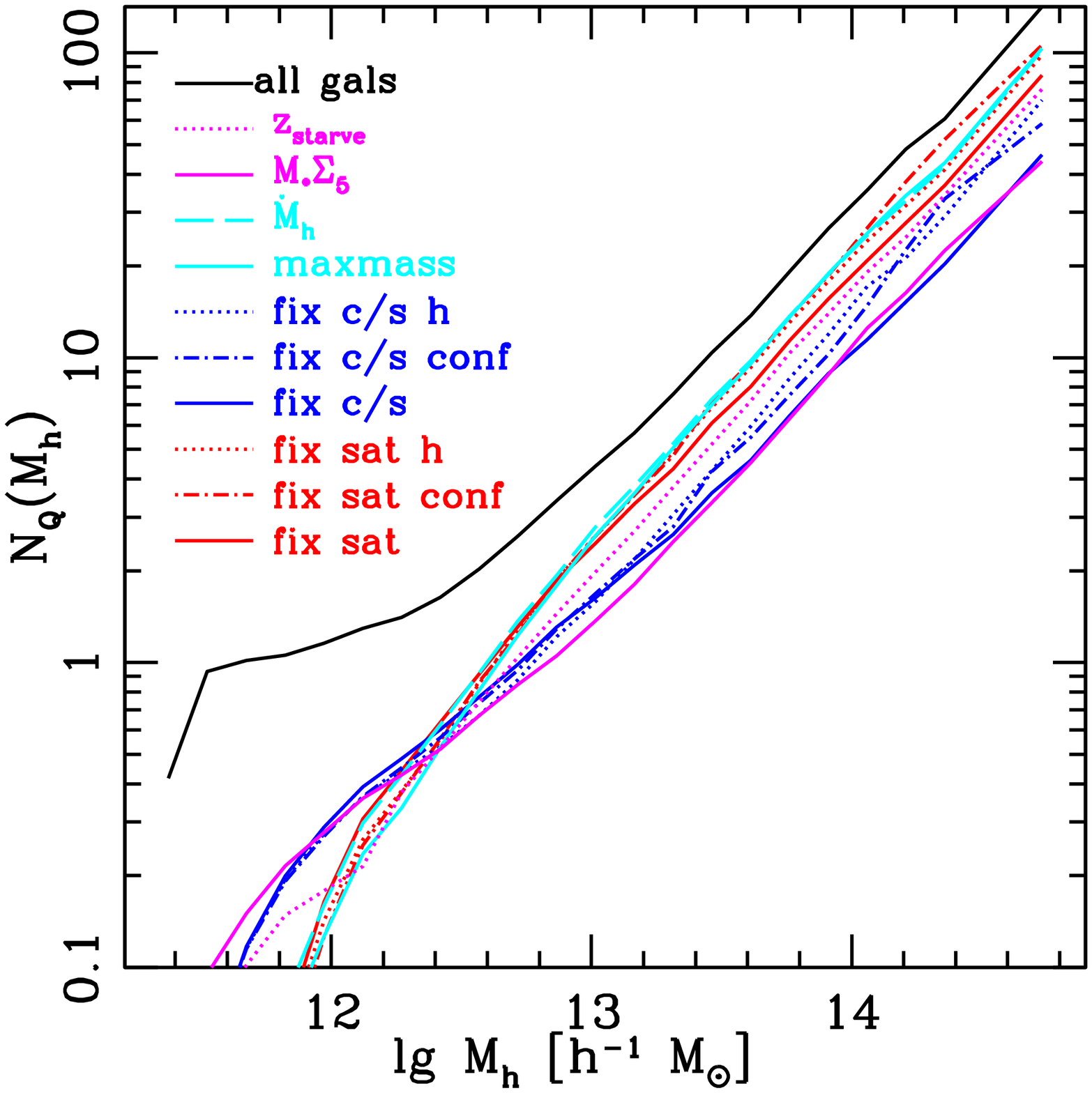}}
\resizebox{3.in}{!}{\includegraphics{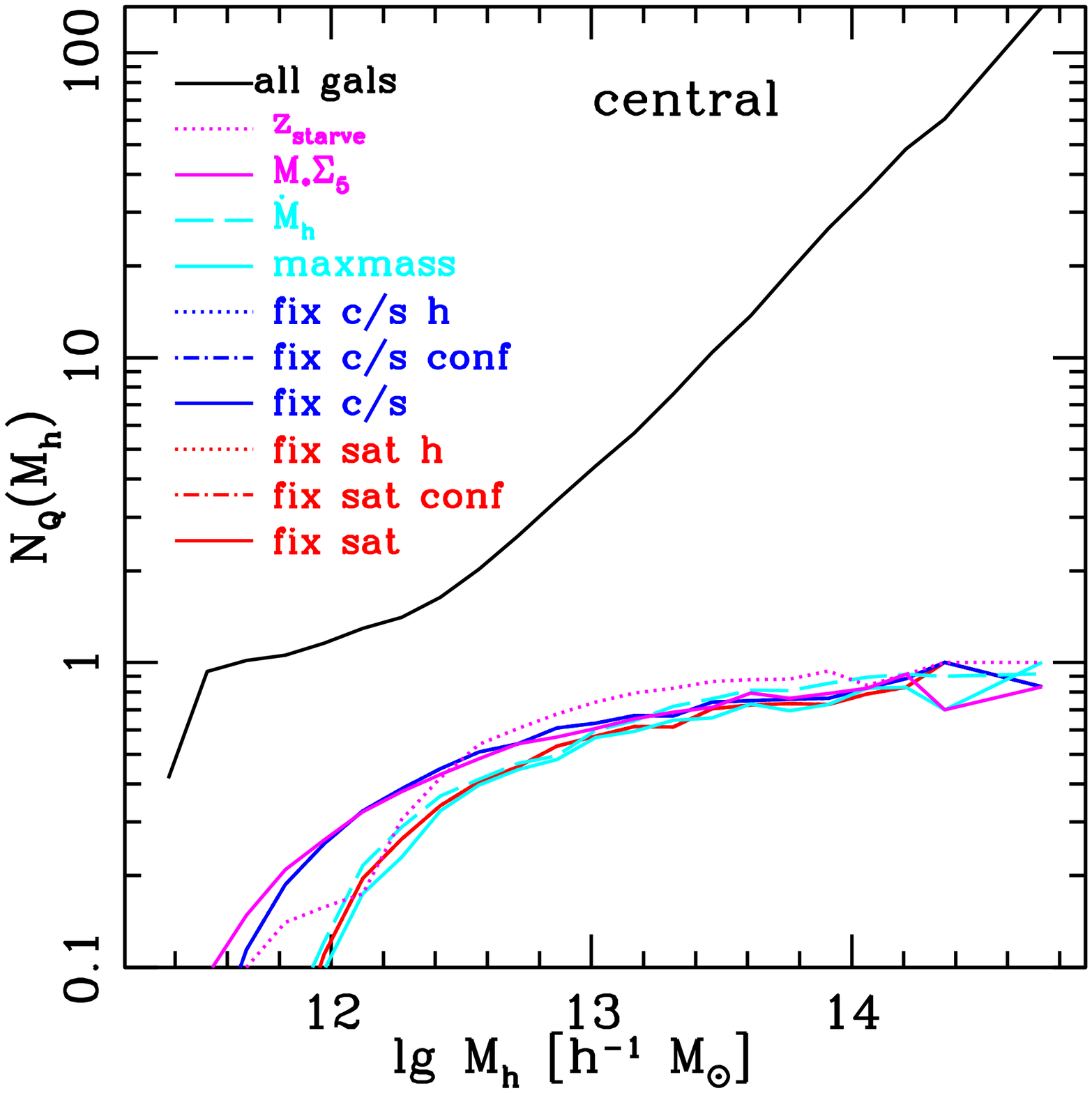}}
\end{center}
\caption{Left: quiescent galaxy HON for our models
  (line types as in Fig.~\ref{fig:redfrac}).
  The top mass bin is wider than the lower ones in order to have at
  least ten halos in it: the halo to halo scatter can be large, especially
  in the ``-conf'' model where changing a central galaxy between quiescent
  and active in principle changes all the satellites as well.
  This leads to the features in the dot-dashed, blue ``fix c/s-conf'' model.
  Note that the quiescent galaxy richness can vary by a factor of two
  between models for the highest mass halos, i.e.~galaxy clusters.
  Right: quiescent central galaxy HON. 
  In both panels, the black solid line is the HON of all galaxies.
}
\label{fig:redhod}
\end{figure*}

Fig.~\ref{fig:redhod} shows the average number of quiescent galaxies per
halo (the ``halo occupation number'' or HON;
\citealt{PeaSmi00,Sel00,CooShe02}), as a function of halo mass.
At left we plot all quiescent galaxies, at right only central quiescent
galaxies.
The average number of all galaxies per halo is also shown for reference as
the black line (and is the same for all models).
The quiescent fraction as a function of halo mass is just the ratio of
each model line to this black line, it is largest at high mass for the
``fix sat-conf'', ``maxmass'' and $\dot{M}_h$ models.  (The active galaxy HON -- the difference of the black solid line and any
of the other curves -- follows a rising power law for all of the
models as well\footnote{We thank P.~Behroozi for discussing this measurement
with us.}).

As was the case for the quiescent fractions, the quiescent galaxy HON's
split roughly into two groupings, which cross around
$M_h\simeq 2 \times 10^{12}\,h^{-1}M_\odot$.

The ``fix c/s'' model has more quiescent, central galaxies at low
$M_\star$ than the ``fix sat'' model.  This in turn implies more
quiescent galaxies in low mass halos and a shallower HON.  Within the
``fix sat'' and ``fix c/s'' classes, the ``-conf'' models have more
quiescent satellites in more massive halos, as these are the halos
most likely to host quiescent centrals.  Similarly, the ``-h'' models
have more quiescent galaxies in massive halos than their random
counterparts because such halos typically host more satellites (of a
given stellar mass) that fell in long ago -- arising from a
combination of reduced dynamical friction and earlier halo formation
time.  A similar effect operates for the
``maxmass'' prescription, because satellites which have reached their
highest mass long ago tend to have fallen in long ago, and thus are again more prevalent in high mass halos.

The $M_\star\,\Sigma_5$ model HON again seems very
similar to the random ``fix c/s'' model.   In both, the satellite and
central quiescent fractions are much closer to each other 
than the other models. 
This large central quiescent fraction at low $M_\star$ can be seen in
Fig.~\ref{fig:satcenred} for both models.  Since low $M_\star$ centrals
live in low mass halos, this in turn explains the low $M_h$ behavior of
the HON.  Again, this behavior seems to be built into the $M_\star\,\Sigma_5$
model by requiring the form of Eq.~(\ref{eq:pengstart}) to have low
enough quiescent fraction at low $M_\star$ (to match our \citet{Dro09}
constraint).\footnote{ Even though quiescence tends
to go with high density and satellites tend to
have higher densities, there are many more low density centrals
at low $M_\star$, which somewhat cancel out the preference for
quiescent satellites.}

It is notable that the quiescent galaxy richness of clusters can vary by
over a factor of two, i.e.~the quiescent galaxy HON is very model
dependent.  This implies that the HON, or the cluster luminosity function
for clusters of a wide range of masses, can provide a strong discriminant
between models (this has been considered in e.g.~\citealt{Tin12}).
The differences in the number of quiescent galaxies as a function of 
host halo mass has implications for the physical role halo environments
and membership play in turning galaxies from active to quiescent.

\subsection{Quiescent Galaxy Distribution in Clusters} \label{sec:clus3d}

In addition to measuring the number of galaxies in a halo of a given mass,
we can ask how these galaxies are distributed within the halos
(i.e.~the profile).
Aside from the $M_\star\,\Sigma_5$ prescription, and to some extent the
``-conf'' models, spatial properties are not used to define which galaxies
are quiescent -- so any dependence arises due to correlations with other
properties which {\it are\/} used in the models.

\begin{figure}
\begin{center}
\resizebox{3.5in}{!}{\includegraphics{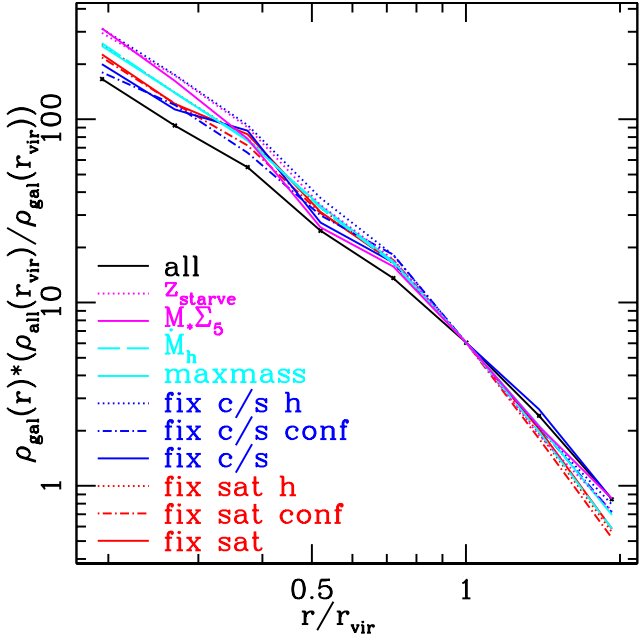}}
\end{center}
\caption{Quiescent galaxy radial profile, for the 18 $z\simeq 0.5$ clusters
  with $M \geq 2\times 10^{14}\,h^{-1}M_\odot$, rescaled at $r=r_{\rm vir}$
  to agree with the total galaxy distribution (black line).
  Line types are as in Fig.~\ref{fig:redfrac}.}
\label{fig:clusprof3d}
\end{figure}

For the 18 most massive clusters in the simulation
($M_h\geq 2\times 10^{14}\,h^{-1}M_\odot$)
we stacked the counts of quiescent galaxies in bins of radial distance from
the most bound particle within the halo (which is very close to the minimum
of the halo potential and the density peak).
We used logarithmically spaced bins in $r/r_{\rm vir}$, where
$r_{\rm vir}$ is the virial radius\footnote{We take
$r_{\rm vir} = 0.92\, r_{180b}$, where $r_{180b}$ is the radius at which
the average density within the cluster reaches 180 times the average
background density.  This radius is measured from the halo center using all
of the particles in the simulation (not just group members).
The factor $0.92$ is the conversion between $r_{180b}$ and $r_{\rm vir}$
for a $c=4$ \citet{NFW} halo at $z=0.5$, calculated as described in
\citet{Whi01}.}. 
The total number of quiescent galaxies in clusters changes from model to
model, as already shown in Fig.~\ref{fig:redhod}.  To highlight the additional
information given by the profile, we normalize the stacked counts for each model
at $r=r_{\rm vir}$.
The resulting profiles are shown in Fig.~\ref{fig:clusprof3d}.

In all of the models the quiescent galaxy radial profiles tend to be more
centrally concentrated than those of all galaxies (the solid black line),
reminiscent of the well-known morphology- and color-density relations
\citep[e.g.][]{Dre97,Tre03,Bal04,Kau04,Pos05,ChrZab05,Loh08,Ski09,Cib13,LacGun13,Muz13}.
Amongst the models, the ``fix c/s-h'' and $z_{\rm starve}$ models are the
most centrally concentrated (this increase in concentration is due to a
quicker drop off in quiescent galaxy density at high radius compared to many
of the other models).
The weak but non-zero correlation between infall time and radius
\citep{OmaHudBeh13}\footnote{Also, Wetzel et al to appear.} %
imprints a radial dependence for the $z_{\rm starve}$ model.
The large number of quiescent galaxies in the $M_\star\,\Sigma_5$ model
near the cluster centers is a direct consequence of the increase of the
quiescent fraction with projected density (Eq.~\ref{eq:pengstart}).

\section{Comparisons: observational} \label{sec:comparison_obs}

In summary, the models produce different numbers of quiescent centrals and
satellites, place them in different halos (HON) and in different
places in halos (profile).
These results are very useful for understanding the differences between the
models but are challenging to access observationally.
They do however drive differences in quantities which can be probed
observationally, for example the change in the HON leads to changes in the
galaxy correlation function or in the richness distribution of groups.
We now turn to such statistics and describe how the differences we have seen
above translate into differences in the observables.

\subsection{Quiescent galaxy correlation function}

One of the most basic measurements that can be made on any population of
objects is its two-point correlation function.  In recent years it has
become standard to use measurements of the correlation function to infer
the halo occupation distribution (shown in Fig.~\ref{fig:redhod}).
We thus expect the correlation function to be a useful diagnostic of these
models.

\subsubsection{Auto-correlations}

\begin{figure}
\begin{center}
\resizebox{3.5in}{!}{\includegraphics{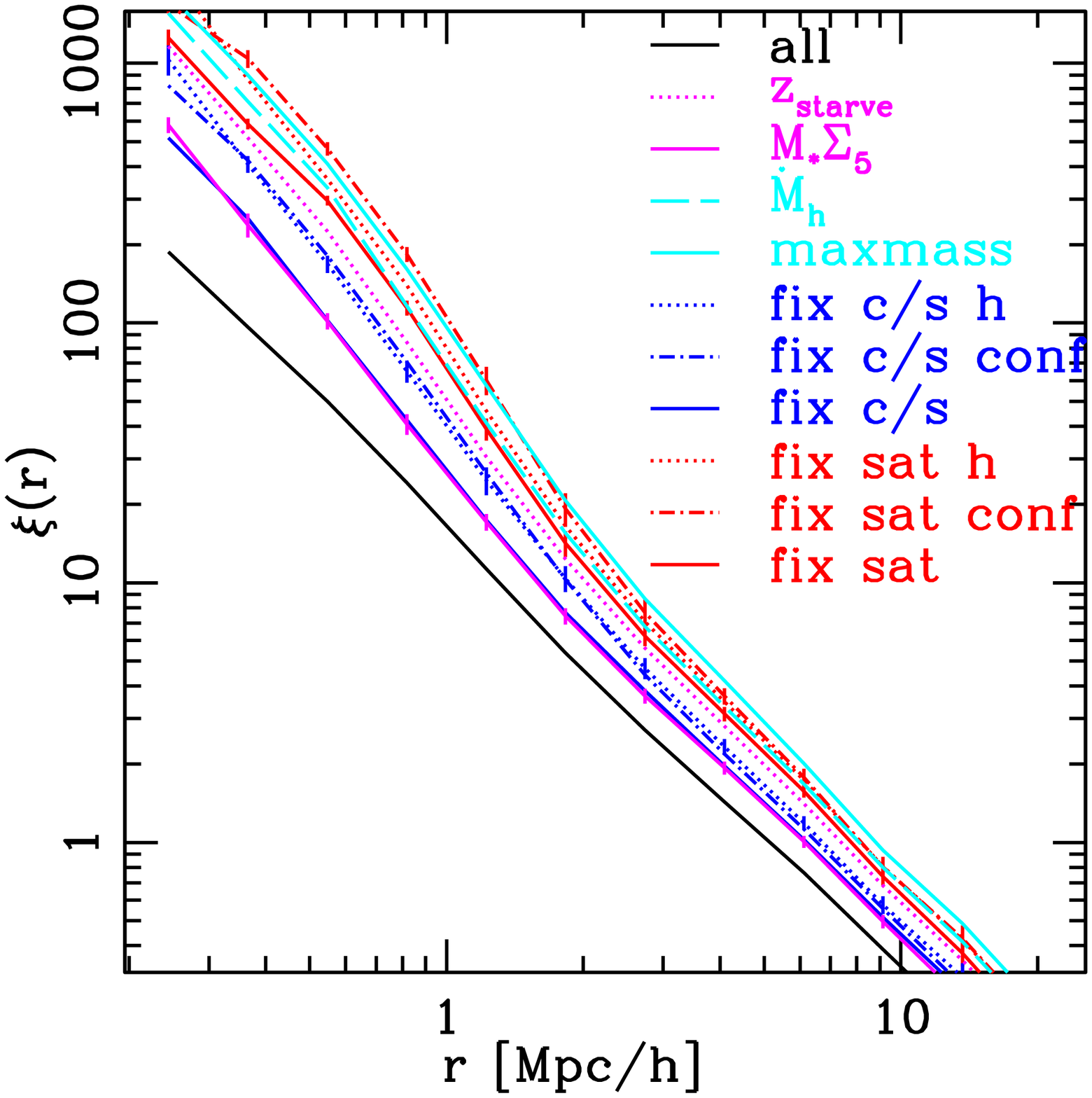}}
\resizebox{3.5in}{!}{\includegraphics{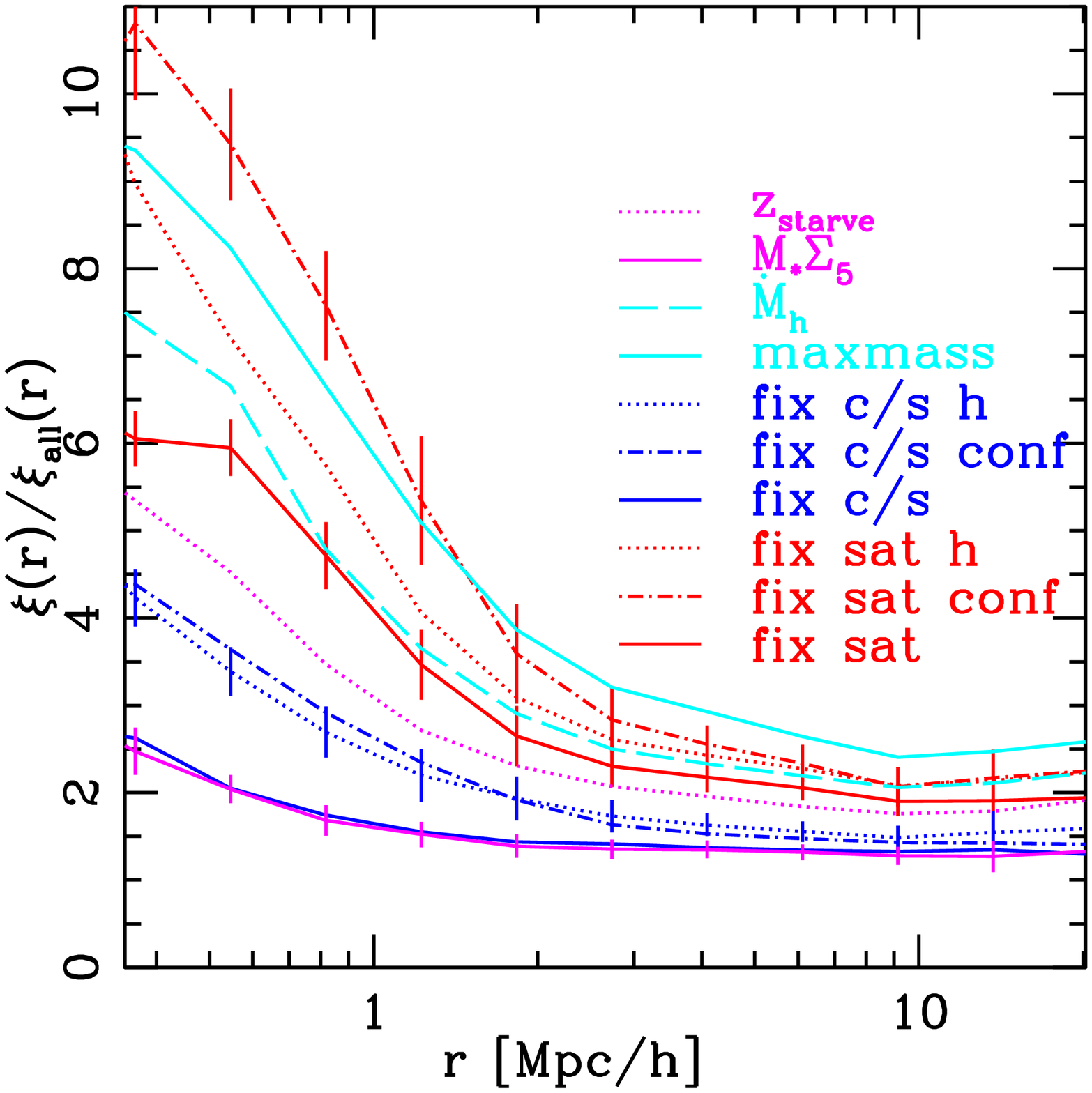}}
\end{center}
\caption{Top: three dimensional real-space correlation function, $\xi(r)$.
  The solid black line is the correlation function of all galaxies,
  $\xi_{\rm all}(r)$;
other line types are as in Fig.~\ref{fig:redfrac}.
  For clarity, representative error bars are shown only for four examples.
  Just as in the HON, the $M_\star\Sigma_5$ model and random ``fix c/s''
  models are roughly degenerate, as are the $\dot{M}_h$ and ``hist sat''
  models (which have $\sim 80\%$ overlap in their populations).  
  The strongest clustered prescription at short distances (red dot-dashed line)
  is ``fix sat conf'', i.e.~the galactic conformity model with the larger
  number of quiescent satellites, as expected again from the HON.
  The $z_{\rm starve}$ quiescent galaxy correlation lies between the other
  two sets of trends, again as expected from the HON in
  Fig.~\ref{fig:redhod}. 
Bottom: $\xi(r)/\xi_{\rm all}(r)$ for each quiescence prescription,
which highlights differences between the models.
Note the linear vertical scale.
}
\label{fig:xi3d}
\end{figure}

\begin{figure}
\begin{center}
\resizebox{3.5in}{!}{\includegraphics{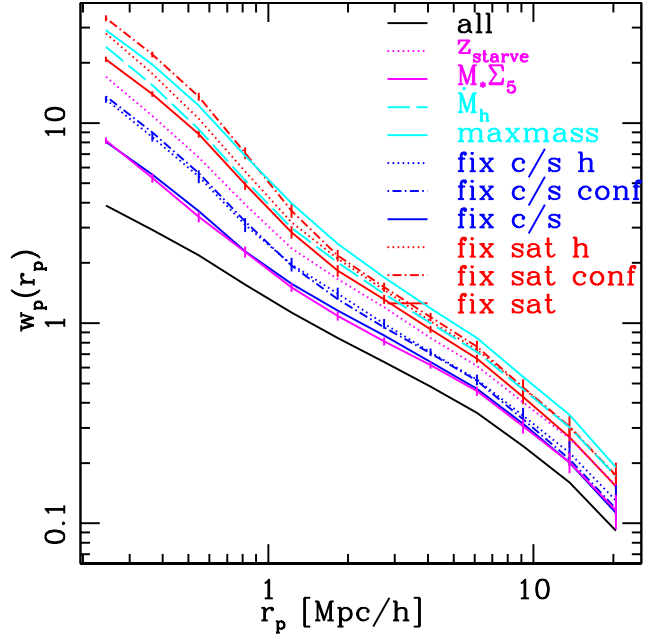}}
\end{center}
\caption{Projected correlation function, $w_p$, for same 10 models as in
  Fig.~\ref{fig:redfrac}, integrating over $\pm 25\,h^{-1}$Mpc in
  the redshift direction.
  The model separations are similar to those seen in the isotropic three
  dimensional correlation function (Fig.~\ref{fig:xi3d}, top).
}
\label{fig:wprp}
\end{figure}

We use the \citet{LanSza93} estimator to measure the (auto-)correlation
functions for our catalogs.
The results for the real-space correlation function are shown at the
top of
Fig.~\ref{fig:xi3d}, with the ratios of each model to the full galaxy
real-space
correlation function below (to highlight the
differences between models).
Errors are shown for only 4 cases to avoid clutter in this and
subsequent figures.  The box is split into octants to compute the
errors, which represent the error in the mean of the octants. They
thus indicate how well a survey of comparable volume to our box could
determine $\xi(r)$.\footnote{While we primarily show real space auto-, cross- and
marked correlations here and below, the trends we focus on are possible from
combinations of their redshift space counterparts such as Fig.~\ref{fig:wprp}, as well as being
accessible via construction of real space quantities themselves.}
Clearly, several of our prescriptions cleanly separate with a volume
comparable to our simulation.  The difference between the curves in 
comparison
to the errors can be used to indicate the observational requirements to
distinguish them.

The correlation function trends are as expected from the halo occupations
in Fig.~\ref{fig:redhod}: models with more galaxies in higher mass halos
cluster more strongly on both large and small scales.
The large-scale clustering is set by the bias, which is determined by the
mean, galaxy-weighted, halo mass.  The small-scale clustering is set by the
number of central-satellite and satellite-satellite pairs within a single
halo.
Satellites with the longest time since infall
(taken as quiescent for ``fix sat-h'')
or since maximum mass
(taken as quiescent for ``maxmass'')
reside in the most massive clusters, so the corresponding quiescent samples
cluster more strongly than the randomly chosen quiescent galaxy sample does.
In addition, the ``-conf'' models tend to put more quiescent satellites in
high mass halos, as most of the high mass halo central galaxies are quiescent.
(This model also has the largest variance as changing one central galaxy to
or from quiescent in a rare halo can significantly change the number of
quiescent satellites for that halo mass and thus the number of quiescent
galaxies with many nearby neighbors.  This suggests that surveys aimed at
ruling out this model need to sample a large volume with a representative
sample of high-mass halos.)

The impact of the radial profile within massive halos on $\xi(r)$ is not
very strong.  While the models separate in Fig.~\ref{fig:clusprof3d} we see
two degeneracies seen in the HON in $\xi(r)$ as well:
the $M_\star\,\Sigma_5$ and random``fix c/s'' models, and $\dot{M}_h$
and ``fix sat-h'' models are also close (the latter two have significant
overlap in their populations as mentioned earlier).  

In practice the three dimensional real-space correlation function is found
by ``deprojecting'' either the projected correlation function
(to minimize the influence of redshift-space distortions) or the angular
correlation function (if only coarse distance estimates are available).
If accurate redshifts are available for the galaxies, the redshift space
correlation function can in principle carry additional information, at
least in part because the small-scale clustering is affected by fingers-of-god
and thus is sensitive to the satellite fraction in massive halos.
To bring the comparison closer to the observational plane, the projected
correlation function
\begin{equation}
  w_p(r_p) = \int_{-z_{\rm cut}}^{z_{\rm cut}} \xi(r_p,z)\ dz \; 
\label{eq:wprp}
\end{equation} 
is shown in Fig.~\ref{fig:wprp}, where we see it exhibits similar behavior
to $\xi(r)$.
In Eq.~(\ref{eq:wprp}), $r_p$ is the separation in the plane of the sky,
$z$ is the separation along the line-of-sight in redshift space and
$z_{\rm cut}$ is a cutoff in the line-of-sight direction.
Ideally $z_{\rm cut}$ is very large so that the the cutoff doesn't matter,
however, our $125\,h^{-1}$Mpc on-a-side octants are relatively small which
argues against using very large $z_{\rm cut}$.
One can still define and measure $w_p(r_p)$ with a small $z_{\rm cut}$, but
the $z_{\rm cut}$ should be matched when comparing theory and observation.
Throughout we use $z_{\rm cut}=25\,h^{-1}$Mpc, thus considering cylinders
$50\,h^{-1}$Mpc deep in the redshift (velocity plus position) direction within
each of the eight $125\,h^{-1}$Mpc octants in the box.

We also computed the angle-averaged, or monopole, redshift correlation
function, $\xi(s)$.  The trends and differences between the models in
$\xi(s)$ closely followed those seen in $\xi(r)$, indicating that both
statistics are comparable in their discriminating power and the additional
sensitivity to e.g.~the fingers-of-god is small.
(The most noticeable difference in $\xi(s)$ is that the ``fix c/s-conf''
model increases its relative clustering at low separation, making it easier
to differentiate it from the ``fix c/s-h'' model.) 

\subsubsection{Cross-correlations}

\begin{figure}
\begin{center}
\resizebox{3.5in}{!}{\includegraphics{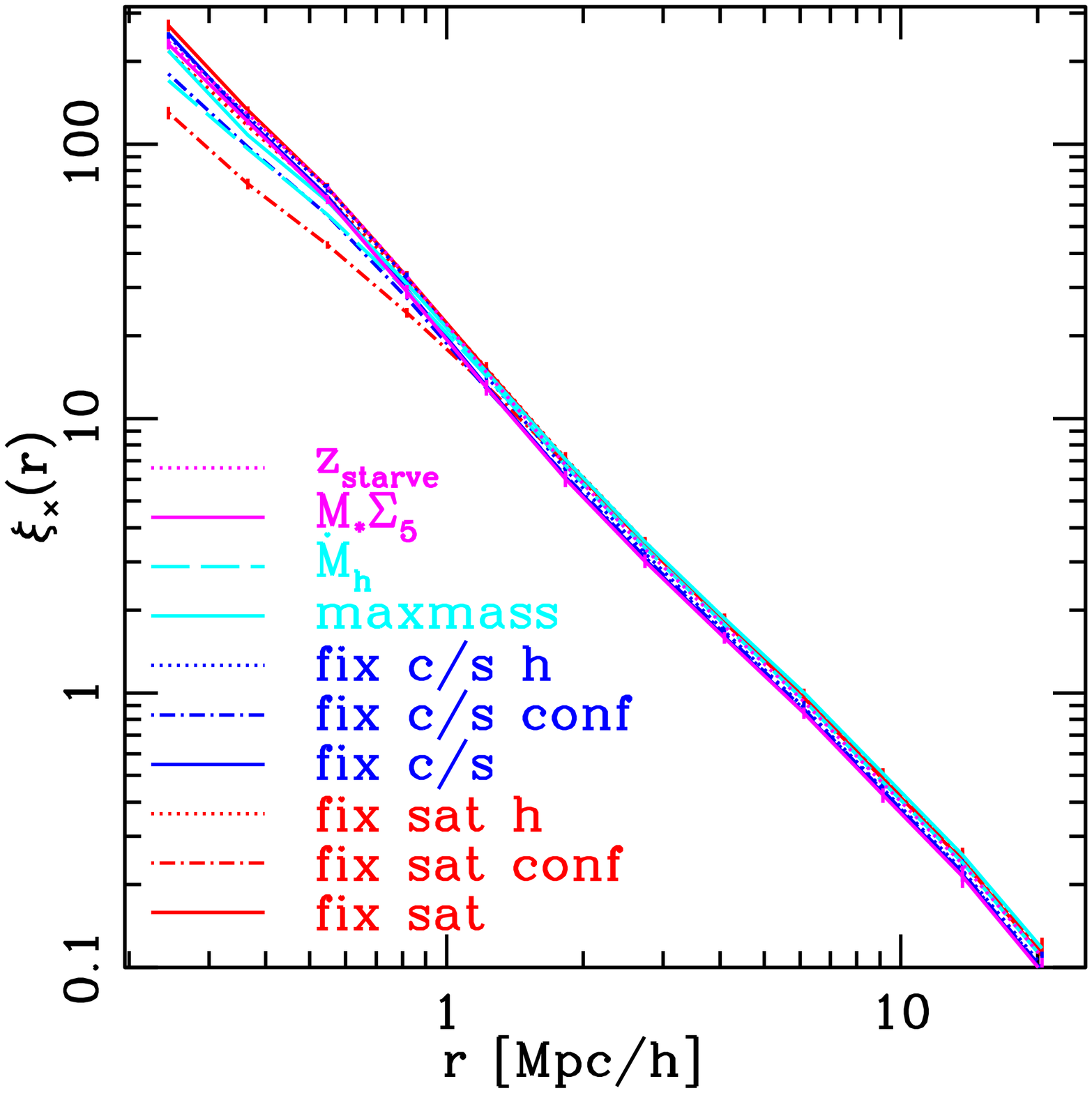}}
\resizebox{3.5in}{!}{\includegraphics{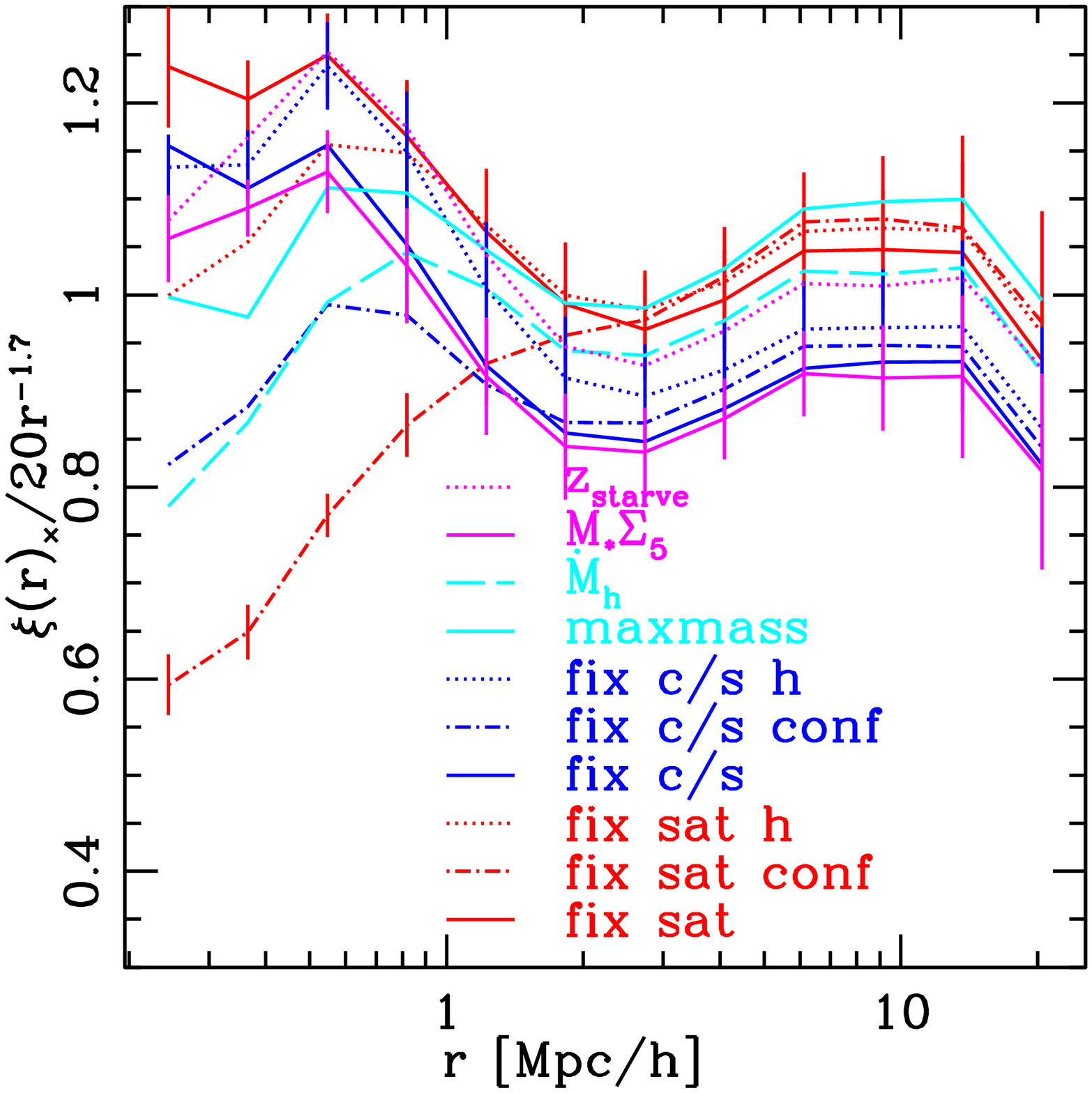}}
\end{center}
\caption{Top:  Three dimensional cross-correlation function between
quiescent and active galaxies for each catalogue.
Line types are as in Fig.~\ref{fig:redfrac}.  The ``fix sat-conf''
model most clearly separates out from the others at small separations.
Bottom:  Dividing by a reference power law model, $\xi_{\rm ref}(r)=
 20 (r[Mpc/h])^{-1.7}$.  Just as in the auto-correlation function,
this highlights differences between
the models, although due to the larger error bars many differences
here
are not significant.
}
\label{fig:xicross} 
\end{figure}

While it is not as commonly employed, it is also possible to measure the
cross-correlation\footnote{We thank F.~van den Bosch for suggesting this
measurement.} between quiescent and star-forming galaxies (or between
either population and the full sample).
As an example we show the cross-correlation
function, $\xi_\times(r)$, between quiescent and star-forming galaxies
at top in
Fig.~\ref{fig:xicross}.
The differences between models are not large, and appear mostly at small 
scales.  
This cross-correlation is most useful in testing the models which invoke
conformity (especially the ``fix sat-conf'' model from the rest).
Models which exhibit galactic conformity have fewer pairs of quiescent and
star-forming galaxies within a single halo (preferring to have all of one
type, the same as the central) so the cross-correlation is suppressed on
small scales.
The trend is less pronounced in the ``fix c/s-conf'' model, which has more
halos where central and satellite galaxy quiescence are mismatched.
The cross correlation divided by a reference power law
model, shown in the lower half of Fig.~\ref{fig:xicross}, more clearly
separates
the models.

\begin{figure}
\begin{center}
\resizebox{3.5in}{!}{\includegraphics{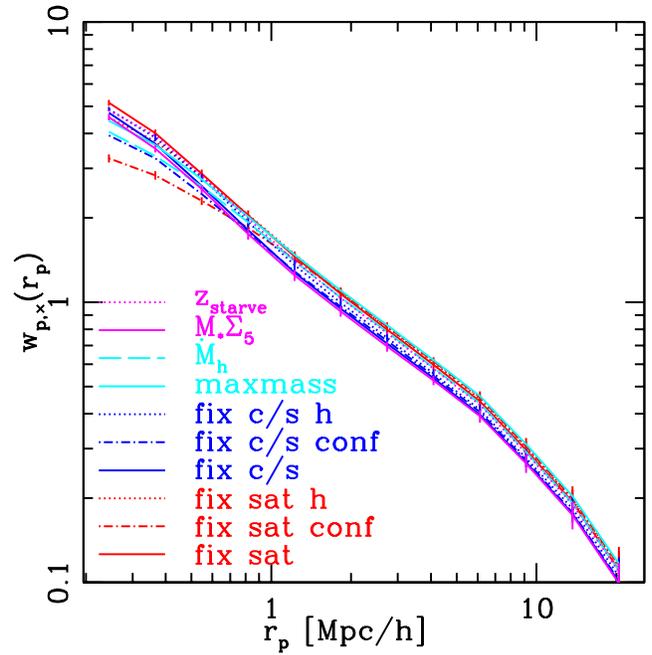}}
\end{center}
\caption{Projected ($\pm 25\,h^{-1}$Mpc in redshift space) 
cross-correlation function between
quiescent and active galaxies for each catalogue, showing similar 
trends to its isotropic counterpart 
at top in  Fig.~\ref{fig:xicross}.
Line types are as in Fig.~\ref{fig:redfrac}.   
}
\label{fig:wpcross} 
\end{figure}

We also show the projected cross correlation function in Fig.~\ref{fig:wpcross}.
It is defined analagously to Eq.~\ref{eq:wprp} and again limited to small
$z_{\rm cut}=25\,h^{-1}$Mpc in the redshift direction because of our box size.

\clearpage
\subsection{Group multiplicity function}

One way to constrain the manner in which galaxies occupy halos is to model
the galaxy correlation function, assuming a particular profile and
parameterized halo occupation distribution.  An alternative method is to
measure the number of groups of objects of a given richness, i.e.~the group
multiplicity function.
Differences between models or between models and observations are of course
easier to interpret the more closely the richness measure (or other observable)
tracks halo mass, but even relatively coarse measures can be interpreted with
the aid of mock catalogs.

\begin{figure}
\begin{center}
\resizebox{3.5in}{!}{\includegraphics{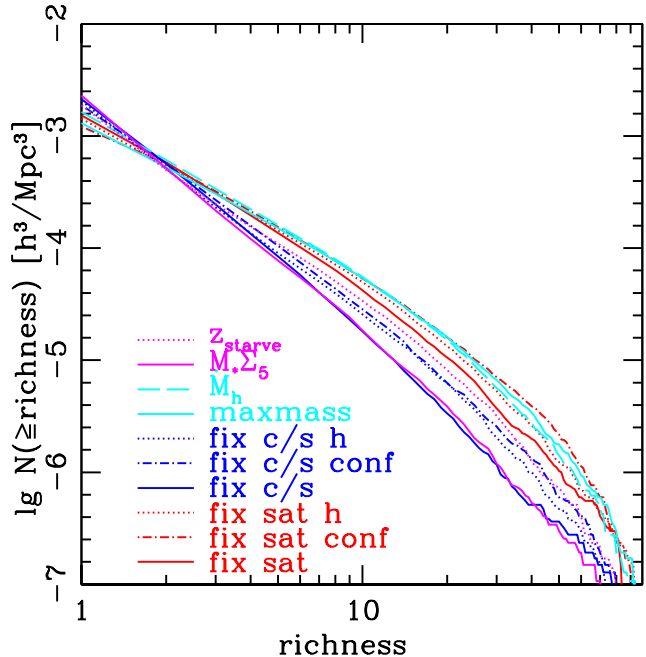}}
\end{center}
\caption{Group multiplicity function (the number density of groups with
$N$ or more quiescent galaxies) for the mocks with line types as in
Fig.~\ref{fig:redfrac}.  Groups are defined using a redshift space
Friends-of-Friends algorithm as discussed in the text.
}
\label{fig:ngtrich}
\end{figure}

There are numerous methods for constructing group catalogs, which rely on a
wide array of differing assumptions.  It is not our intention here to perform
an exhaustive comparison, but rather to illustrate the potential of this
measurement as a diagnostic.
For our example we assume we have redshifts for each of our quiescent galaxies
and find groups using a Friends-of-Friends algorithm in redshift space
\citep[e.g.][]{HG82}.
We follow \citet[][see also \citealt{Hea13a}, who used it to study abundance
matching]{Ber06} and set the linking lengths in the line-of-sight and
perpendicular directions as $b_\perp=0.14$ and $b_\parallel = 0.75$, both
measured in units of the mean inter-galaxy separation and including only
quiescent galaxies in the input catalog.
We define richness to be the number of (quiescent) galaxies associated with
each group.
Fig.~\ref{fig:ngtrich} shows the cumulative number density of groups as a
function of richness, for the different catalogs.

The basic trends in Fig.~\ref{fig:ngtrich} can be understood by reference
to the HON in Fig.~\ref{fig:redhod}.
The ``fix c/s'' and $M_\star\,\Sigma_5$ models have the fewest rich groups,
because they have the fewest quiescent galaxies per massive halo
(Fig.~\ref{fig:redhod}).
By contrast the ``maxmass'' and ``fix sat-conf'' models, which have many
quiescent galaxies in massive halos, have a relatively large number density
of rich groups.  In general we see that the imperfections involved in
constructing the group finder (i.e.~the impurity and incompleteness) do not
qualitatively change the trends present in the underlying models.
This suggests that a sufficiently accurate group catalog would provide a
strong discriminant between the models.  Because it opens up a large parameter
space, we leave an examination of other group finding methods, and methods
which can work with photometric redshifts, for future work.

\subsection{Two dimensional profile in clusters}

One can also consider the observational counterpart of the three
dimensional cluster profile (Fig.~\ref{fig:clusprof3d}), the two
dimensional cluster profile. Here galaxies are counted within a
redshift cylinder of the cluster center.  However, accurately
estimating the observational scatter is more challenging with our mock
catalogues, as it relies upon the accuracy of the cluster mass
(determining $r_{\rm vir}$) and cluster centering, both of which
include significant assumptions besides those used in the construction
of the catalogue.
To bound the information one could obtain from this measurement we consider
the case where the group center and $r_{\rm vir}$ are known perfectly.
In this limit the most notable distinction this measurement provides
is between the $M_\star\,\Sigma_5$ and ``fix c/s'' models.
However as we will see below, this can be obtained with other more
direct observations.

\subsection{Distribution of projected density: $\Sigma_5$}

The $M_\star\,\Sigma_5$ model of \citet{Pen10} used a projected density,
$\Sigma_5$, in determining which galaxies are classified as quiescent.
This is a property that can be used to help discriminate amongst models,
even when the models make no explicit reference to it.
There are many types of density one could define (for a comparison,
see, e.g.,
\citet{HaaSchJee12,Mul12}),
we shall consider $\Sigma_5$ because it has been studied in this
context by \citet{Pen10} and is used as input to one of our models.

\begin{figure}
\begin{center}
\resizebox{3.in}{!}{\includegraphics{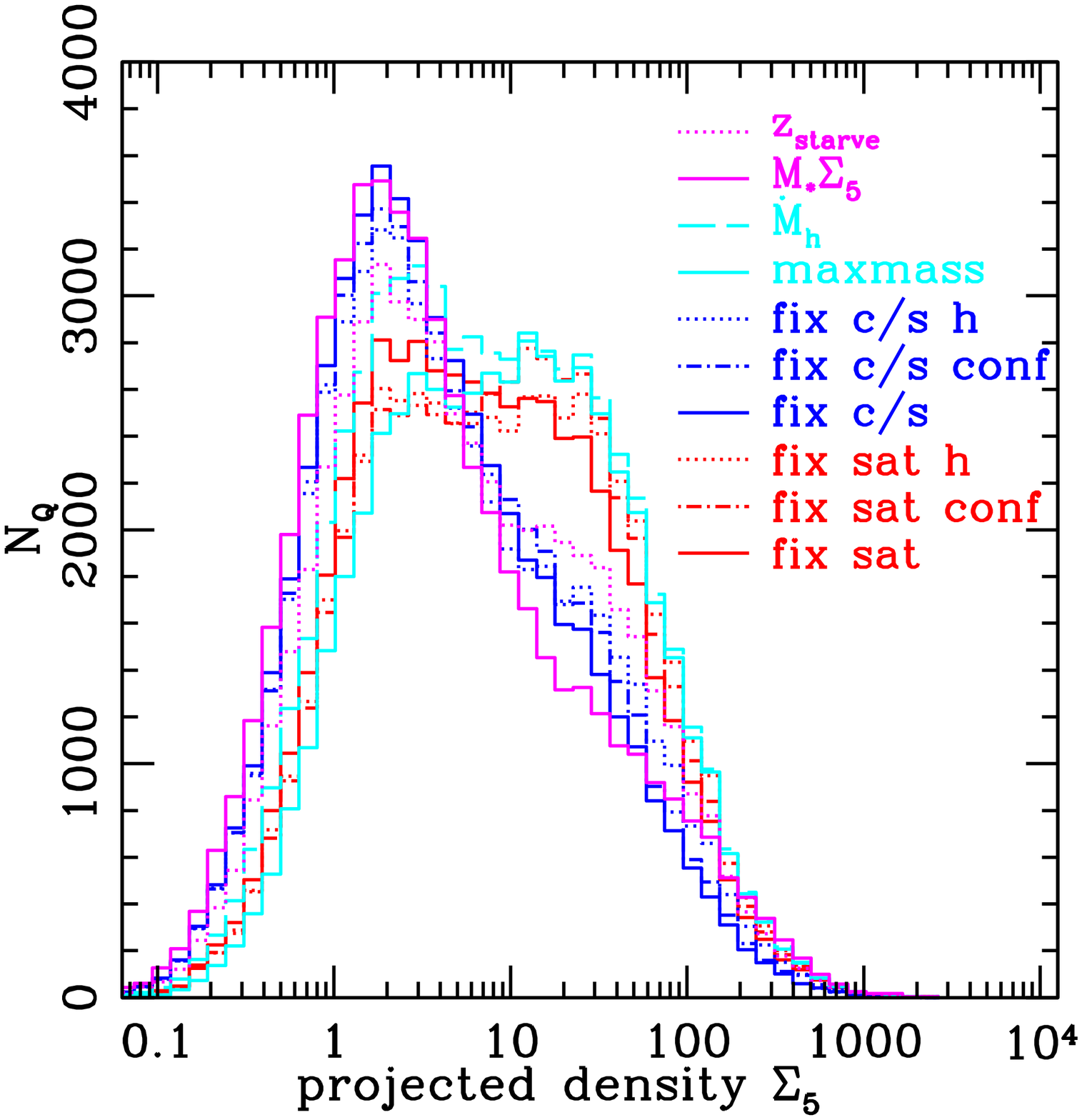}}
\resizebox{3.in}{!}{\includegraphics{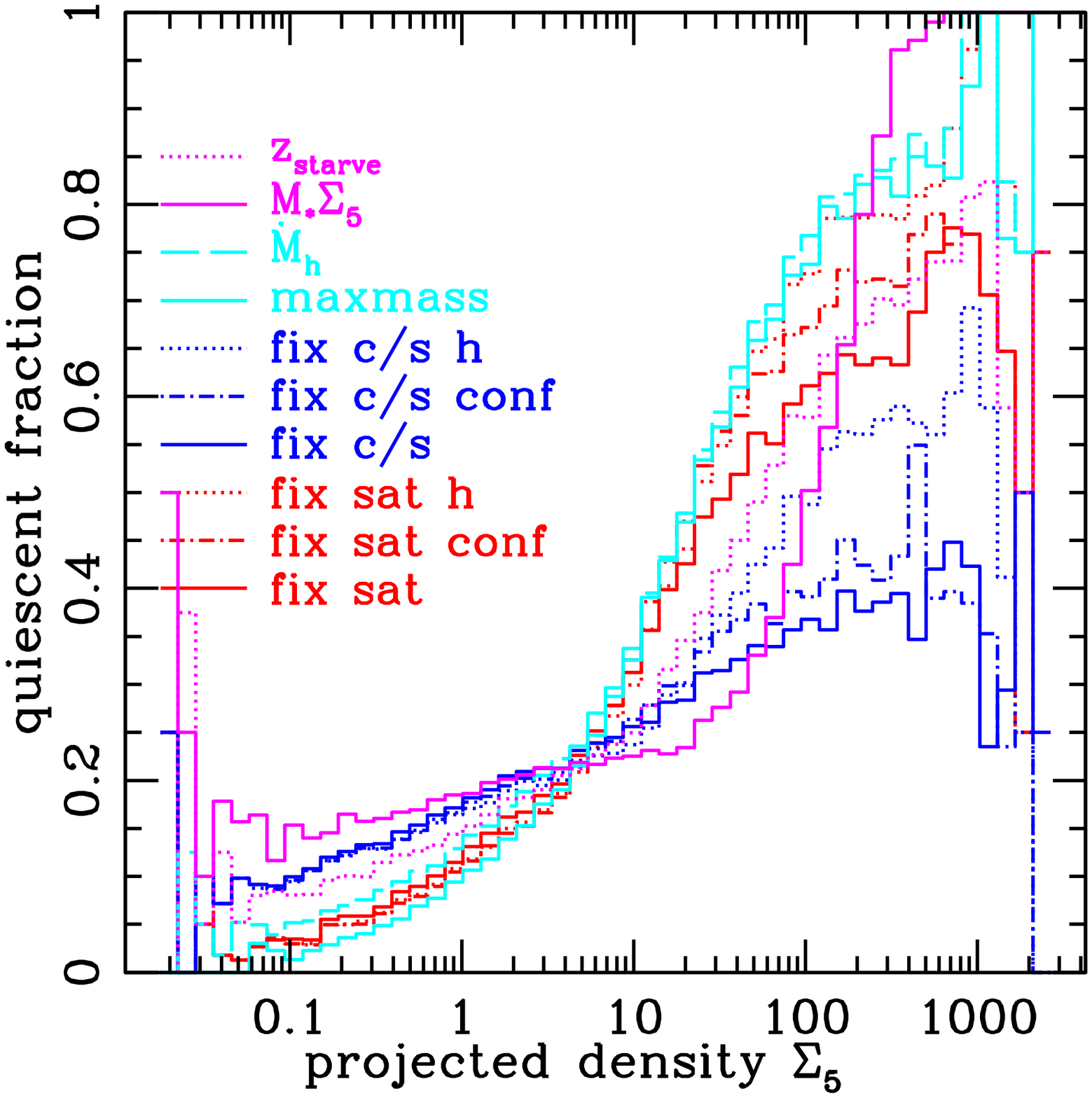}}
\end{center}
\caption{Total number (top) and fraction (bottom) of quiescent galaxies at a given projected density, $\Sigma_5$,
which are quiescent.  Line types as in Fig.~\ref{fig:redfrac}.}
\label{fig:dens}
\end{figure}

The distribution of $\Sigma_5$ across the whole galaxy sample is approximately
lognormal, as expected.  However the distributions of $\Sigma_5$ for the
quiescent galaxies, in raw counts and as a fraction of all galaxies
(Fig.~\ref{fig:dens}), discriminate quite strongly among the models. 
Models split into two groups, mostly determined by the quiescent satellite
fraction as a function of stellar mass.
Models with a large number of quiescent central galaxies tend to have overall
more galaxies at lower $\Sigma_5$ (Fig.~\ref{fig:dens}, top).
(Central galaxies are a minority population for high densities, only
reaching at least
 half of the galaxies for projected densities $\la 10$.)
A slightly different division between the models is seen in the quiescent
galaxy fraction (Fig.~\ref{fig:dens}, bottom).   Differences become the most
pronounced at the high density tail ($\Sigma_5\ga 10^2$), composed almost
entirely of satellites in rich groups or clusters.
The general shape of the quiescent fraction (Fig.~\ref{fig:dens}, bottom)
is expected: the rise of the quiescent fraction to high $M_\star$ implies
that higher $M_\star$ satellites are generally quiescent and such satellites
lie predominantly in massive halos.

The ``fix c/s'' model has the lowest quiescent fraction
at high $\Sigma_5$ -- Eq.~(\ref{eq:fixcensat}) requires many low-$M_\star$,
central galaxies to be quiescent.  This means the satellite galaxies in the
same $M_\star$ bin must be star-forming, removing the high-$\Sigma_5$
population in Fig.~\ref{fig:dens}.  The ``fix c/s-conf'' model favors
galaxies in high mass halos with a star-forming central to be
star-forming,
removing quiescent galaxies from these high density halos.  This is somewhat
counteracted
for the ``fix c/s-h'' models as the longest lived satellites tend to
be in the higher mass halos.
The $M_\star\,\Sigma_5$ model has the highest quiescent fraction for highest
$\Sigma_5$, partly by construction.  Next highest are the ``fix sat-h'',
$\dot{M}_h$ and  ``maxmass'' models.
These models turn galaxies quiescent based upon infall time and galaxies that
fell in earlier tend to be in higher mass halos and thus denser environments.

In is encouraging that two models which were quite close in their HON,
$M_\star\,\Sigma_5$ and ``fix c/s'', separate cleanly by looking at the
high projected density tail of the quenched fraction.

\subsection{Marked Correlation Functions}

In addition to looking at the distribution of $\Sigma_5$, we can use
the projected density as a weight when computing the two point function,
creating a ``marked'' correlation function \citep{SheConSki05,Har06,Ski13}.
Such a function is straightforward to compute if $\xi(r)$ can be computed
and can be used to break degeneracies in modeling the occupation distribution
\citep{WhiPad09}.  

The marked correlation function, $M(r)$, is defined as
\begin{equation}
  M(r) = \frac{1}{n(r) \bar{m}^2} \sum_{ij} m_i m_j
\label{eq:markedcorr}
\end{equation}
where the sum is over all galaxy pairs, $i$ and $j$, which are separated
by $r$ and $m_i$ and $m_j$ are the marks, $n(r)$ is the number of pairs in
the bin and $\bar{m}$ is the average mark in the sample.
$M(r)\neq 1$ on scales where the clustering of objects depends upon the mark.
One of the advantages of a marked correlation function, from an observational
perspective, is that it is simple to modify a code which computes $\xi(r)$ to
also compute $M(r)$.  No further information is needed (beyond the marks and
galaxy positions).  In fact, one does not even need a random catalogue due to
the cancellation between the numerator and denominator.
It is also relatively easy to estimate the errors \citep{SheConSki05}.
We will focus here on the three dimensional marked correlation function but
there is no reason one cannot use the projected or angular version instead.

\begin{figure}
\begin{center}
\resizebox{3.5in}{!}{\includegraphics{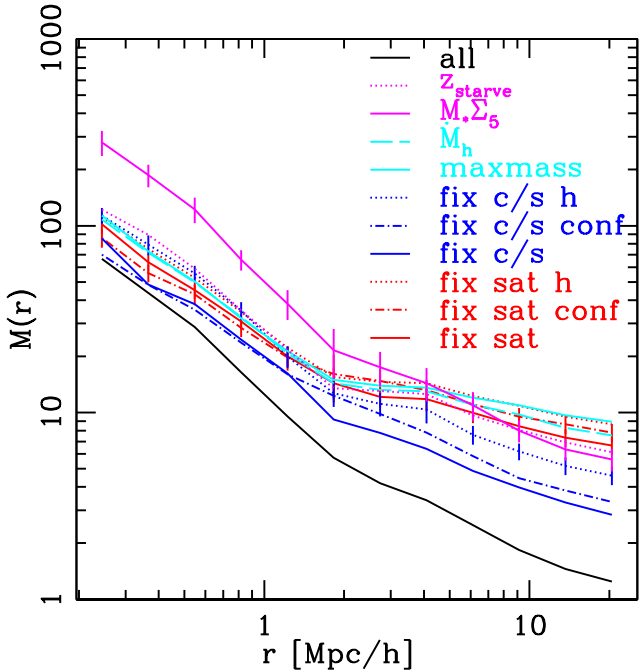}}
\end{center}
\caption{Marked correlation function, $M(r)$, for quiescent galaxies.
  Line types are as in Fig.~\ref{fig:redfrac} and the solid black line
  is $M(r)$ for all galaxies.
  The mark is the $5^{\rm th}$ nearest neighbor projected density,
  $\Sigma_5$, of \citet{Pen10}.
  Error bars indicate the error on the mean $M(r)$ from 8 disjoint octants
  of the box.  The $M_\star\,\Sigma_5$ models separates out cleanly
  from the others.}
\label{fig:ximark}
\end{figure}

If we take our mark, $m_i$, to be the projected density $\Sigma_5$ of the
$i$th galaxy, the resulting $M(r)$ is shown in Fig.~\ref{fig:ximark}.
For comparison, we also show the marked correlation function for all
(not just quiescent) galaxies as the solid black line. 
As expected, the $M_\star\,\Sigma_5$ quiescent galaxies cluster differently,
as the mark $\Sigma_5$ was used in their selection.
There is a feature between $1$ and $2\,h^{-1}$Mpc, roughly at the virial
radius of massive halos.
Some similarities between the model ordering at small radius and the stacked
and rescaled cluster profiles (Fig.~\ref{fig:clusprof3d}) is apparent.

\begin{figure}
\begin{center}
\resizebox{3.5in}{!}{\includegraphics{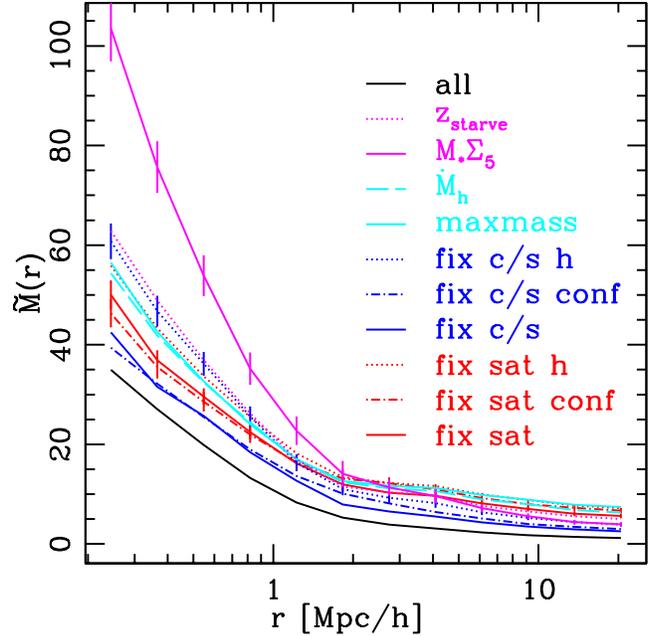}}
\end{center}
\caption{Marked correlation function, $\widetilde{M}(r)$, for quiescent
galaxies using Eq.~(\ref{eqn:sigmaprime}) as the mark with
$\Sigma_{\rm max}=300$.
Line types are as in Fig.~\ref{fig:ximark} and the solid black line
is $\widetilde{M}(r)$ for all galaxies.}
\label{fig:ximarkprime}
\end{figure}

If the dynamic range in the mark is very large, then there is some concern
that $M(r)$ can become sensitive to rare outliers.  A simple way around this
is to modify the mark from $\Sigma_5$ to
\begin{equation}
  \widetilde{\Sigma} \equiv
    \frac{\Sigma_{\rm max}\ \Sigma_5}{\Sigma_{\rm max}+\Sigma_5}
\label{eqn:sigmaprime}
\end{equation}
with $\widetilde{\Sigma}\simeq\Sigma_5$ for $\Sigma_5\ll\Sigma_{\rm max}$ and
$\widetilde{\Sigma}\le\Sigma_{\rm max}$.  Fig.~\ref{fig:dens} suggests that
$\Sigma_{\rm max}=300$ is a reasonable choice, and we take that as our 
fiducial value.  The resulting $\widetilde{M}(r)$ is shown in 
Fig.~\ref{fig:ximarkprime}.
Similar behavior is seen as we vary $\Sigma_{\rm max}$ from 30 to 300
suggesting the features we see in Fig.~\ref{fig:ximark} are stable and
robustly present in the catalogs.
As $M(r)$ and $\widetilde{M}(r)$ combine projected density and galaxy number,
an enhancement at small scales could be due to many galaxies with a small
mark, or few galaxies with a large mark.  To differentiate, we removed the
3-7 per cent of galaxies with $\Sigma_5\ge 10^2$ from the catalog.
This only decreases the radius below which these models dominate
(to $\sim 800\,h^{-1}$kpc/h), but does not erase the separation.
Thus we infer the small-scale enhancement is driven by a large number of
galaxies, not simply the dense tail of the distribution.

In this section we considered several observational measurements and
how they separated out the different quiescent galaxy catalogues.  Now
we turn to general results and implications of the intrinsic and
observational measurements the the ensemble of catalogues.

\section{Discussion} \label{sec:discussion}

We generated a set of mock galaxies by assigning stellar masses to subhalos
in an $N$-body simulation via abundance matching.  These mock galaxies match,
by construction, the stellar mass function of \citet{Dro09}, as shown in
Fig.~\ref{fig:smf}.  We then marked a fraction of these galaxies as quiescent,
using 10 prescriptions adapted from the literature.  All of the prescriptions
are tuned give a good fit to the total quiescent fraction as a function of
$M_\star$, as shown in Fig.~\ref{fig:redfrac}, but differ in their assumptions
about which properties are important in determining quiescence and in how they
incorporate centrality, environment and history.  

The number of
implied quenching mechanisms differs as well.  Two are
used in the
``fix sat'', ``fix sat-h'',``fix c/s'', ``fix
c/s-h'' models (for satellite and for central galaxies), and for the
${M_\star\,\Sigma_5}$ model (stellar mass and projected density),
while ``fix sat-conf'', ``fix c/s-conf'' might allow one
 mechanism across a shared halo.  The ``maxmass'', $\dot{M}_h$ models
have one mechanism, combined with no mass gains for satellites, while the
$z_{\rm starve}$ model has three mechanisms.

We investigated the consequences of these different quiescence assumptions
for intrinsic galaxy properties and observational measurements, at one fixed
redshift.
Models differed in intrinsic properties such as halo occupation and galaxy
profiles within massive halos, although the central/satellite split as a
function of stellar mass was degenerate for several of them.
Observationally, we considered measurements derived from galaxy positions
and stellar masses.  Promisingly, we found that almost all of the models
could be separated from each other by some combination of observations, if
measurement errors could be made small enough. 
While many of the models could be distinguished based on very traditional
measurements such as the auto- and cross-correlation functions, we also saw
that less often considered statistics such as the group multiplicity function
and the (projected) density-marked correlation function allowed us to further
break degeneracies between models.
We advocate that these statistics be reported in future.

We now turn to intrinsic and observational properties and how they
relate to each other and to mechanisms/models for quenching star
formation.  We saw that the fraction of ${\rm lg}M_\star/M_\odot\simeq 10$
quenched galaxies which were central or satellite roughly divided the
models into two groups.  The strongest division was between different
choices for the evolution of the quenched satellite fraction with
redshift, as expected, but this division also strongly separated the
$M_\star\,\Sigma_5$ model from the ones in which ${\rm SFR}\propto\dot{M}_h$
or where quiescence was based upon the time at which the subhalo achieved
its maximum mass, for example.
This division further implied a change in the way quiescent galaxies
occupied halos, i.e.~the HON (Fig.~\ref{fig:redhod}) which showed up
clearly in the relative clustering (Fig.~\ref{fig:xi3d}).  This argues
that measures of quiescent galaxy clustering can be used to constrain
the redshift evolution of the quiescent fraction of satellite and
central galaxies and thus the timescale over which star-formation
quenching acts \citep[see also][for similar conclusions]{AJCF,Tin13}.

Models which are degenerate in their satellite or central
quenched galaxy fractions as a function of stellar mass can still
differ significantly in other measures, depending upon environment and
history.  Those which invoke conformity tend to have the largest 
number of quiescent satellites in massive halos within their class.  
Observationally this leads to enhanced auto-correlation, since massive
halos have a large bias.
These models also have a depressed cross-correlation between active and
quiescent galaxies at small separations, as galaxies in a halo tend to
all be either active or quiescent.  
Conformity might point to quenching properties acting on larger scales than
those of a halo, as for fixed stellar mass it can give similar
quenching likelihoods to satellites at the
outskirts and satellites near
the center.

The models based upon conformity have central galaxy assignments identical 
to those of their associated models based on satellite infall or random
selection.  It was therefore encouraging that there were significant,
observable differences associated with the variations in satellite quiescence
mechanisms.
(This is in part why conformity was introduced at lower redshifts in the
 SDSS \citep{RosBru09}, but here it also separated out the models where
 satellite infall was related to quiescence.)

The associated ``-h'' models are best further considered along with the other three
models using infall time (where satellite infall stops mass gain or generally
increases quiescence probability -- our ``maxmass'', $\dot{M}_h$ and
$z_{\rm starve}$ models).
If halo growth is considered a proxy for baryonic accretion (and satellite
infall a proxy for strangulation, or cessation of baryonic accretion), these
can be thought of as ``feeding limited'' models.
Observationally, these five models most prominently group together in the
projected density marked correlation functions $M(r)$ and $\widetilde{M}(r)$.
They are large at short distances, only being surpassed by the
$M_\star\,\Sigma_5$ model (which uses $\Sigma_5$ in its construction).
This observation thus seems to separate out models which are feeding limited
or in which group pre-processing played an important role.

The differences between the three ``feeding limited'' models which
have similar satellite populations (``fix sat-h'', ``maxmass'',
$\dot{M}_h$) are interesting as well.  All quench satellites after
infall (with similar although not identical decay times).  The first
assigns the central galaxies randomly, while the other two depend upon
halo mass gain or time of maximum mass.  Differences between these
models might thus indicate which intrinsic properties and observations
are impacted by the history of central mass gains.  (Although the
central and satellite quiescent fractions are not identical, in raw
numbers there are more differences in the quiescent central
populations.)  They are quite close in the full HON at high mass, but
differ very slightly at low stellar mass (also seen in the central
galaxy HON).
The three models separate out most clearly (but not by much) from each
other in the auto-correlation function.  The ``maxmass'' is slightly
larger at high richness in the group
multiplicity function, and the $\dot{M}_h$ model also has a relatively
depressed quiescent-active galaxy cross-correlation at short
distances.  Otherwise, the history of central mass gains
seems to not have much of an effect on the observations we considered.

All of the models make an explicit distinction between central and satellite
galaxies except for the $M_\star\,\Sigma_5$ model.  Due to the way it is
constructed, this model shows a very different distribution of $\Sigma_5$
for quiescent galaxies than the other models, which results in a strong
enhancement of the projected density-marked correlation function at
small scales, and the quiescent fraction as a function of
projected density.  In the other observations it is close to
degenerate with the model sharing its HON and profile, 
which assigns quiescent galaxies depending only upon centrality
(``fix c/s'').  

Several observations were quite good proxies for the intrinsic differences
we saw between the models.  For example, the group multiplicity function we
used orders models in almost the same way as the HON at high richness or
halo mass (only two models reverse).
The two dimensional stacked cluster profile has information similar to
the three dimensional stacked cluster profile and the projected density
distribution (for $\Sigma_5\sim 10-100$) showed a similar ranking of models
to the quiescent satellite fraction at intermediate $M_\star$.
If one can reliably infer which galaxies belong to which dark matter halo from
a group catalog, and has a reliable method of determining the central galaxy,
then the central and satellite quiescent fractions become directly observable.
As we saw, these functions were extremely helpful in partitioning the models.
Conversely the differences in quiescent fraction at high projected density
and differences in the radial profile may impact the calculations of purity
and completeness for color-selected cluster finders which are trained
on mock catalogs (which make either implicit or explicit assumptions about these
properties).

We used ten catalogues with quiescence based upon dark matter subhalo
histories, environment and centrality.  However, these tests could be
expected to shed light on many other proposed models, including
those based upon fewer simulation properties (e.g. without subhalos, such as
\citealt{PeaSmi00,Sel00,CooShe02}), more simulation properties
(e.g. including histories, for instance,
 \citealt{AJCF,MosNaaWhi13,MutCroPoo13,Cat13,Bau06,Ben12}, or baryons, e.g., \citealt{OWLS,Nyx,Vog13}), or other simulation properties (e.g., density
as in \citealt{Sco13}).  
Other tests may also add more information
(e.g. lensing was used by \citet{MasLinYos13} for similar tests at
lower redshifts, we did not consider it here as galaxy shapes are
required as well).  Some tests to constrain quiescence were recently reported by
\citet{Tin13} at similar redshifts (clustering, weak lensing, group
catalogue) as this work was being prepared for publication.

\section{Conclusions} \label{sec:conclude}

One of the most striking features of the galaxy population is that it
exhibits bimodality in color, morphology and star-formation rate.
In particular, the existence of two broad types of galaxies
(those which are actively star-forming and those which are quiescent)
cries out for a theoretical explanation.
Unfortunately the range of physics and of physical scales involved in
setting the star-formation rate in a galaxy is enormous, making empirical
models of the phenomenon of great importance.
Many such models have been proposed to explain the color or star-formation
rate bimodality at $z\simeq 0$.
Motivated by the wealth of data on the higher $z$ Universe which we expect
to have soon from large surveys, and using high-resolution cosmological
$N$-body simulations, we have investigated the predictions for a range of
models for star-formation quenching to see in what way they differ and what
observations (of what accuracy) can be used to discriminate amongst them.
The advantage of using large surveys is that statistical errors can be
controlled and are more accessible and one can disentangle the many
correlations between galaxy properties which limit inferences from small
samples.  The disadvantage is that information is more circumstantial and
so requires a different sort of detective work.

We have focused our attention at $z\simeq 0.5$, which is at high enough
$z$ that we expect significant galaxy evolution but low enough $z$ that we
expect to have large statistical samples of galaxies in the near future.
Note that $z\simeq 0.5$ is approximately $5\,$ Gyr ago, which is longer than
the main-sequence lifetime of $>1.4\,M_\odot$ stars.
Any galaxy less massive than $2\times 10^{10}\,M_\odot$ on the star forming
main sequence would at least double its stellar mass in this period.
It is also around $z\simeq 0.5$ that we see a rapid rise in the number
density of intermediate-mass, quiescent galaxies, making this a particularly
interesting time to study.

We considered several different galaxy properties which are expected to
correlate with star-formation quenching and investigated how statistical
measurements might be used to distinguish them.
We created different mock catalogues, all sharing the same stellar mass
for each galaxy and close to the same overall quenched fraction as a
function of stellar mass, but with different criteria for classifying a
galaxy as quiescent.  
Many of these models had degeneracies in basic intrinsic properties
(satellite/central fraction, for instance), or halo occupation (HON),
but the suite of observations we used could nonetheless separate them out,
given small enough measurement errors.
Fixing the quenched central galaxies and varying the satellite quenching
gave models which were distinguishable, similarly, models where the majority
of differences were in the central galaxies also could be differentiated.

More specifically, in addition to the quiescent galaxy auto and
cross-correlation functions and quiescent fraction as a function of
stellar mass, we found that projected density counts, the projected
density marked correlation function, and the group multiplicity function
could further serve to separate models, and also seemed to correlate with
centrality in host halo, feeding limited quenching and halo occupation
respectively.
These observational measurements increase the galaxy formation information 
available from surveys possessing galaxy stellar masses and 
positions. 

We have mostly considered how these observational measurements can be used to
distinguish models from each other, using observations.
Alternatively this provides a particularly discriminating set of tests for
validating mock catalogues, constructed by any means.  They can also
be used to decode associations of other galaxy properties with
galaxy histories, environments and centrality, for instance AGN
activity or morphology.

While we were completing this work we became aware of \citet{Tin13} which 
compared several different measurements of a survey at similar
redshifts to constrain quiescent fraction evolution and of \citet{Hea13b}
which considered measurements at low redshift to constrain color and developed
a formalism for comparing dark matter simulation properties which are
correlated with color. 

\bigskip

We thank E. Bell, F. van den Bosch, K. Bundy, N. Dalal, C. Knobel,
M. George, A. Hearin, C. Heymans, C. Lackner, Z. Lu, P. Behroozi,
D. Watson, A. Wetzel for conversations, and D. Watson, A. Hearin,
R. Skibba, A. Wetzel and the referee for helpful suggestions on
the draft.
JDC also thanks the Royal Observatory, Edinburgh, and Kavli IPMU for
their hospitality and
invitations to speak on this work as it was being completed.
JDC was supported in part by DOE.
MW was supported in part by NASA.

\end{document}